%% file: main.tex
\documentclass[letterpaper,twocolumn,10pt]{article}
\usepackage{usenix2019_v3}

\usepackage{graphicx}
\usepackage{subcaption}

\usepackage{booktabs}
\usepackage{multirow}
\usepackage{tablefootnote}
\usepackage{array}
\newcolumntype{C}{>{\centering\arraybackslash}m}
\usepackage{xurl}
\usepackage{amsfonts,amsmath, amssymb}
\usepackage{pifont}
\usepackage[]{mdframed}
\usepackage{algpseudocode}

\makeatletter
\def\blfootnote{\gdef\@thefnmark{}\@footnotetext}
\makeatother

\input{commands}

\begin{document}

\renewcommand{\sectionautorefname}{Section}
\renewcommand{\subsectionautorefname}{Section}

\date{}

\title{\Large \bf Nebula: Self-Attention for Dynamic Malware Analysis}

\author{
{\rm Dmitrijs Trizna}\\
{\rm dtrizna@microsoft.com}\\
{\rm Microsoft Corporation, Czech Republic}\\
{\rm University of Genova, Italy}
\and
{\rm Luca Demetrio}\\
{\rm luca.demetrio@unige.it}\\
{\rm University of Genova and Pluribus
One, Italy}\\
\and
{\rm Battista Biggio}\\
{\rm battista.biggio@unica.it}\\
{\rm University of Cagliari and Pluribus
One, Italy}\\
\and
{\rm Fabio Roli}\\
{\rm fabio.roli@unige.it}\\
{\rm University of Genova and Pluribus
One, Italy}\\
} 

\maketitle

\blfootnote{© 2024 IEEE.  Personal use of this material is permitted. Permission
from IEEE must be obtained for all other uses, in any current or
future media, including reprinting/republishing this material for
advertising or promotional purposes, creating new collective works,
for resale or redistribution to servers or lists, or reuse of any
copyrighted component of this work in other works.}

\begin{abstract}
Dynamic analysis enables detecting Windows malware by executing programs in a controlled environment and logging their actions. 
Previous work has proposed training machine learning models, i.e., convolutional and long short-term memory networks, on homogeneous input features like runtime APIs to either detect or classify malware, neglecting other relevant information coming from heterogeneous data like network and file operations. 
To overcome these issues, we introduce Nebula, a versatile, self-attention Transformer-based neural architecture that generalizes across different behavioral representations and formats, combining diverse information from dynamic log reports.
Nebula is composed by several components needed to tokenize, filter, normalize and encode data to feed the transformer architecture. We firstly perform a comprehensive ablation study to evaluate their impact on the performance of the whole system, highlighting which components can be used as-is, and which must be enriched with specific domain knowledge.
We perform extensive experiments on both malware detection and classification tasks, using three datasets acquired from different dynamic analyses platforms, show that, on average, Nebula outperforms state-of-the-art models at low false positive rates, with a peak of 12\% improvement. 
Moreover, we showcase how self-supervised learning pre-training matches the performance of fully-supervised models with only 20\% of training data, and we inspect the output of Nebula through explainable AI techniques, pinpointing how attention is focusing on specific tokens correlated to malicious activities of malware families.
To foster reproducibility, we open-source our findings and models at \url{https://github.com/dtrizna/nebula}.
\end{abstract}

\section{Introduction}

Dynamic malware analysis is a crucial task not only for detecting but also for understanding the threats that are widespread over the entire Internet.
Once samples are collected, analysts execute malware inside isolated environments (sandboxes or emulators), where they list all the actions performed by the program like network and filesystem access, registry modifications, API calls, and kernel syscalls~\cite{siem_survey}.
These actions are then summarized into textual reports, which are manually analyzed by experts to distill the rationale behind the maliciousness of the analyzed sample.
This task is tedious and resource-intensive since it involves domain experts in the process and manual labeling.


Machine learning (ML) techniques, particularly Convolutional Neural Networks (CNNs) and Long Short-term Memory (LSTM) models, are now widely utilized to streamline this process.
These models are trained on vast volumes of textual reports, allowing quicker classification of new inputs and reducing human intervention~\cite{quoVadis, neurlux, cruparamer, edr, empirical_edr, zhang_dmds}. 
While CNNs capture local patterns in reports, providing valuable features for neural architectures, LSTM models learn global token relationships~\cite{lstm, attention_first_paper}. 
However, these proposed schemes are hindered by three main downsides: (i) convolutions only capture local information, discarding the global correlations contained in reports between actions, while LSTM models struggle in modeling sample behavior based on prolonged token sequences, like a chain of API calls with arguments; (ii) most of the proposed techniques solely rely on homogeneous input data, like API calls~\cite{cruparamer,quoVadis,zhang_dmds}, rather than leveraging more complete and heterogeneous information representing the behavior of malware samples; and (iii) source code, data, and pre-trained models are typically not available for most of the proposed techniques, hindering reproducibility.

\begin{figure*}[ht]
    \centering
    \includegraphics[width=\textwidth]{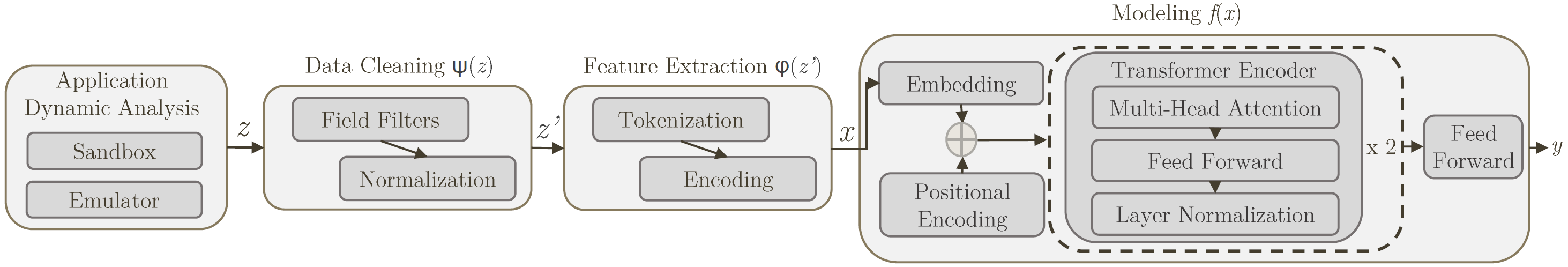}
    \caption{A schematic overview of Nebula.}
    \label{fig:scheme}
\end{figure*}

To overcome these issues, we present Nebula, an ML model based on the Transformer architecture~\cite{transformer_paper} trained on reports of different nature and formats.
Unlike traditional models, Nebula leverages the \textit{self-attention} mechanism inherent in Transformer neural networks, granting Nebula the capability to discern both local and global relationships in a report.
To the best of our knowledge, we are the first to propose general Transformer architecture to tackle both malware detection and classification from raw dynamic log reports.
Instead of solely focusing on few portions of reports, we design Nebula to properly work on all the output provided by sandboxes, thus making Nebula able to correlate tokens from different sources.
To build Nebula, we consider several data cleaning approaches and and feature extractors, and we deeply study their effect through an extensive ablation study (\autoref{sec:ablation}).
Through this analysis, we highlight that some standard NLP techniques, like tokenization through Byte Pair Encoding (BPE) can be applied ``as-is", while it is necessary to preprocess data through the lenses of domain knowledge, by replacing mostly-unique tokens like specific IP addresses, hashes, and internet domains.
We then test Nebula against different state-of-the-art approaches leveraging both CNNs and LSTMs, and we benchmark their performances on both malware detection and classifications tasks span on three distinct datasets acquired from different sandbox environments (\autoref{sec:comparison_sota}).
From our analyses we can conclude that Nebula is, on average, the best model to handle both tasks on all the considered dataset with performances aligned to the state of the art.
In particular, we highlight a peak of improvement of 12\% true-positive rate on a specific dataset with an very low false-positive rate of $10^{-3}$ setting.
This result is achieved under a strict regime of very low false positive rate, which is a crucial aspect for deployed systems~\cite{edr, empirical_edr}.
Also, we inspect how self-supervised learning can reduce the number of training data needed to fine-tune Nebula on the malware detection task (\autoref{sec:ssl}).
Our results exhibit a positive trend, highlighting that Nebula achieves state-of-the-art performance when firstly pre-trained on 80\% of training dataset, and fine-tuned for the downstream task of malware detection with only the remaining 20\% of samples.
Lastly, we review the output of Nebula through the lenses of explainable AI techniques~\cite{sundararajan2017axiomatic,bertviz} (\autoref{sec:explainability}), by confirming that our model focuses on tokens specific to certain malware behaviors, backed up by domain knowledge.
To foster reproducible results, we do not only share the code and pre-trained models of Nebula, but we also re-implement, re-train and release methods that were previously closed source~\cite{zhang_dmds, neurlux}.\footnote{\url{https://github.com/dtrizna/nebula}}


\section{Dynamic Windows Malware Analysis}
\label{sec:background}

This section provides the background information necessary to understand the technical advancements made by Nebula. 

\begin{table*}[t!]
\caption{Dynamic malware analysis modeling techniques.}
\centering
    \input{table1}
\label{tab:models}
\end{table*}

\subsection{Malware Behavioral Reports}
\label{sec:logs}

System compromise can have different manifestations depending on impact, like sensitive data exposure or the misuse of computational resources. Adversaries employ a diverse range of methods, from utilizing built-in tools and protocols aligning with a ``living-off-the-land'' approach to leveraging stolen credentials or employing social engineering tactics to achieve their goals through legitimate user accounts. 
Threat actors often deploy their own software agents, referred to as malware. 
According to the 2022 Verizon Data Breach Investigation Report~\cite{verizondbir}, malware was responsible for nearly 40\% of breaches.
Malware analysis can be segregated into static and dynamic methodologies. The former entails the evaluation of software samples without executing them.
On the contrary, dynamic software analysis is a process that commences with the ``detonation'' of a sample in a controlled environment. 
Dynamic analysis is done by isolation of application, preventing it from impacting other system parts, while maintaining the realism of potential target system to extract malware actions, producing a behavioral report~\cite{malware_experiment_practices}.
These are readable text files that summarize all the meaningful events captured by the sandbox, and they are listed to help the analysis task conducted by humans.

\subsection{Machine Learning Pipeline for Dynamic Analysis}
\label{sec:pipeline}

Machine learning has become a significant element in malware analysis, with efficient modeling schemes proposed for both static and dynamic data structures derived from malware.
We now describe how ML can employ textual reports by introducing three main steps: (i) \textit{data cleaning} to prepare raw data; (ii) \textit{feature extraction} to create a mathematical representation of reports; and (iii) \textit{modeling} the problem to train the final classifier, as depicted in \autoref{fig:scheme} given modeling part is performed by Nebula.

\myparagraph{Data Cleaning.} 
First, the behavioral report is cleaned and normalized to make the data manageable for further processing. Filters are used to remove unnecessary data and preserve only a specific set of fields, while normalization techniques are applied to systematize values that are stochastic in nature and do not correlate with application behavior like hash-sums or IP addresses. This allows us to introduce domain knowledge \cite{usenix_inside_re_head} and, as shown in \autoref{sec:ablation}, improves the model's generalization abilities by reducing variability in values irrelevant to the prediction. We denote this step as $\vct z^\prime = \psi(\vct z)$, where $\vct z$ is the raw data collected from the dynamic analysis environment, and $\vct z^\prime$ is the cleaned and normalized textual data.

\myparagraph{Feature Extraction.} Then, $z^\prime$ undergoes feature extraction denoted $\vct x = \phi(z^\prime)$. As a final step, producing a numerical array $x$, suitable for analysis by ML model. Feature extraction $\phi$ involves a dichotomy between (a) feature engineering or (b) token encoding. Feature engineering involves the manual or automated selection and transformation of relevant features from the cleaned data $z^\prime$, for instance, feature hashing applied to API call names \cite{zhang_dmds} or regular expression-based feature extractors \cite{zhang_dmds}. 
Token encoding involves a tokenization step, which transforms the textual data $z^\prime$ into a sequence of tokens and a vocabulary $V$ of all possible tokens. Tokenization can be based on regular expressions,
be influenced by a domain knowledge \cite{trizna_slp}, or involve statistical methods like Byte-Pair Encoding (BPE) \cite{bpe, sentencepiece}.
The sequence of tokens is then encoded into a numerical array $x$ using an encoding function $f$, which might be as simple as one-hot encoding, be calculated with term frequency-inverse document frequency (TF-IDF), or use embedding function $f: V \rightarrow \mathbb{R}^d$, where $d$ is the embedding dimension.

\myparagraph{Modeling.} The final step is to use the numerical array $x$ as an input to a ML model $f(x)$, which produces a prediction $y$ of a malware label.
The modeling function can be as simple as linear models like logistic regression. However, for behavioral reports, the best schemes incorporate representations of sequential information. This can be achieved by convolutions, recurrent neural networks or self-attention with positional encoding.

\subsection{Review of Dynamic Models}
\label{sec:existing_model_review}

The landscape of behavioral malware analysis showcases a competitive interplay between commercial solutions and academic research, with different attitude towards a modeling an adversary.
Commercial anti-virus (AV) and Endpoint Detection and Response (EDR) products have integrated behavioral analytics into their detection methodologies, forming part of a multi-objective heuristic that leverages both static and dynamic analysis.
The behavioral components of their multi-objective heuristics are closed, which prohibits their disentanglement on the user side for comparison purposes.
This lack of transparency means that we cannot gauge how much of the overall performance of these commercial solutions is attributed to their behavioral modeling component specifically.
Also, AVs and EDRs work in \emph{real-time} settings, implying that decisions are taken in a matter of milliseconds, opposed to sandbox analyses that are conducted \emph{offline}, and later evaluated thoguh reports.
Due to these discrepancies, in this work we will only focus on academic dynamic malware analysis conducted offline through sandbox analyses, since its comparison with AVs and EDRs would be unfair.
%
%
In academic research, we encounter several groundbreaking methodologies in dynamic malware analysis that pose a formidable challenge to the current state-of-the-art. To offer a consolidated view of these promising approaches, we have curated a selection of these solutions in \autoref{tab:models}, systematizing their respective pipelines according to the steps introduced in \autoref{sec:pipeline}.
A common theme among contemporary academic contributions is the employment of traditional techniques, such as one-dimensional convolutions, optionally complemented with recurrent layers through Long Short-Term Memory (LSTM)~\cite{lstm}, as part of their core modeling approach $f$. However, each of these methodologies introduces a unique approach in either data cleaning ($\psi$) or feature extraction ($\phi$) processes, thereby diversifying the analytical landscape of dynamic malware analysis.

\myparagraph{Neurlux (Jindal et al.~\cite{neurlux}).} A distinctive feature of this approach is the absence of operations during the data cleaning phase ($\psi$), passing raw behavioral reports directly to the feature extraction process ($\phi$). This phase involves a simple whitespace tokenization procedure and sequences encoding with a vocabulary size of $V=10,000$. The resulting sequences are then modeled $f$ using a combination of one-dimensional convolutions, LSTM, and conventional attention mechanisms \cite{attention_first_paper}, which is applied to the output of the LSTM layer. Their code is publicly accessible; therefore, we are able to compare our results with this model. However, the data utilized in their research remains undisclosed.

\myparagraph{Gated CNN (Zhang et al.~\cite{zhang_dmds}).} This model introduced an analysis where $\psi$ preserves only API call data, each undergoing a custom feature engineering process during $\phi$ phase. Then, the sequence of featurized vectors is modeled though a gated convolution network as $f$. While the code for their model is not released publicly, it was provided by the researchers upon request, enabling us to draw a direct comparison between our results and their model. However, similar to the case of Neurlux, the data utilized in their study has not been released.

\myparagraph{Quo.Vadis (Trizna~\cite{quoVadis}).} This hybrid model simultaneously assesses contextual, static, and dynamic features. Their model code is released publicly, which significantly contributes to the transparency of their work. For our analysis, we concentrated on the dynamic component of their pipeline, which data cleaning $\psi$ preserves only API call names. Feature extraction $\phi$ label-encodes each API call name with a vocabulary of $V=600$ and subsequently models $f$ with a 1d convolutional neural network. This work is especially notable for its public release of a comprehensive dataset consisting of Speakeasy~\cite{speakeasy} emulation reports. This allows for the pursuance of both malware detection and type classification objectives.

\myparagraph{JSONGrinder (Bosansky et al. \cite{avast_data}).} This model provides a unique method for parsing hierarchical JSON reports, originally proposed in~\cite{jsonGrinder_Mandlik}. This method employs a combination of Julia libraries, specifically \texttt{JsonGrinder.jl} used for feature extraction $\phi$ and \texttt{Mill.jl} for modeling $f$, data cleaning $\psi$ is omitted. The $\phi$ phase infers a Hierarchical Multiple Instance Learning (HMIL) schema from the data, constructing a fixed-size vector, while the modeling is based on a multilayer perceptron (MLP) for sample classification. However, it is worth noting that their implementation was not compatible with the latest version of Julia (v1.8.5) at the time of our experiments, causing the original model implementation to fail without modifications. Additionally, Bosansky et al.'s work is notable for its release of a comprehensive dataset useful for malware family classification, which adds considerable value to the existing body of resources in this field.

\myparagraph{CurParamer (Chen et al.~\cite{cruparamer}).} This method preserves only API calls from the original report during the data cleaning phase ($\phi$). The feature extraction step ($\psi$) involves a unique approach to API labeling and embedding, which includes parameter-assisted API labeling and sensitivity-inspired API embedding. These techniques utilize domain knowledge to generate more efficient numerical representations of API calls. To model these representations ($f$), they employ two separate networks based on 2D convolutions and LSTM. Although their feature extraction methodology is intriguing, it is presented with little implementation details, which reduces its replicability. Despite efforts to access the modeling code, the authors made no public version available, even upon private request.

\section{Nebula: Transformer Architecture for Dynamic Malware Detection}
\label{sec:nebula}

The design of our dynamic malware analysis pipeline draws from the proven success of the attention mechanism in Natural Language Understanding (NLU). Particularly, the self-attention-based Transformer architecture \cite{transformer_paper} has demonstrated superior performance over conventional RNN- or CNN-based modeling methods \cite{gpt, devlin2019bert}. 
These successes guided the selection of techniques used during our feature extraction ($\phi$) and modeling ($f$) stages. The most significant deviation from standard NLU pipelines is evident during the data cleaning phase ($\psi$). Here, we employ a domain-specific parser that (i) retains only those fields from the original structured report relevant for behavior generalization; and (ii) normalizes unconstrained and arbitrary values within such selected fields.
In the feature extraction phase ($\phi$), we tokenize each report into a sequence of tokens of length $N$ and encode each token based on a vocabulary of size $V$. When modeling this sequence, we first embed the input vector to a higher dimension, apply position encoding, and then process it through a Transformer encoder layer to apply the self-attention operation. The resulting attended tensors are then forwarded to a classifier that produces the final prediction. A high-level overview of our Nebula modeling scheme is depicted in \autoref{fig:scheme}.

\subsection{Data Cleaning}
We detail the data cleaning $\vct z^\prime = \psi(\vct z)$ applied by Nebula.

\begin{figure}
    \centering
    \includegraphics[width=0.92\linewidth]{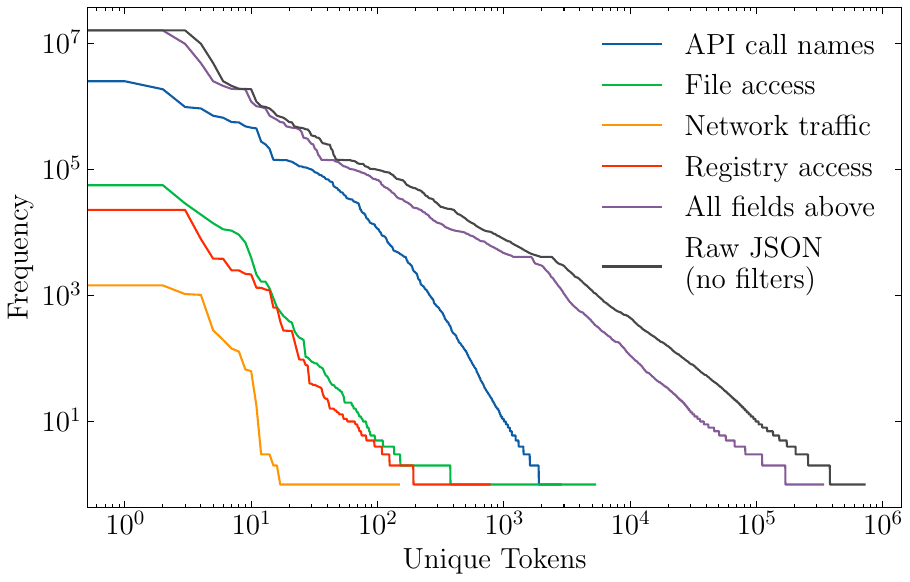}
    \caption{Visualization of whitespace token frequency in the Speakeasy~\cite{speakeasy} emulated behavioral report training set \cite{quoVadis}.}
    \label{fig:token_counts_fields}
\end{figure}

\myparagraph{Vocabulary and Field Filters.} Machine data is more volumetric and heterogeneous than natural languages. Therefore, it can have a significantly larger vocabulary, as no distinct lexical boundaries or grammatical rules define the language being used. 
In system logs, it is common to see arbitrary character combinations like \texttt{/tmp/83afba/setup.bin} or \texttt{jre1.8.0\_311}, which explode vocabulary given improper handling. For instance, even after path normalization, we observe more than 6000 unique filepaths, where only roughly 400 paths repeat, and the rest appear only once.
The \autoref{fig:token_counts_fields} visualizes the frequency distribution of tokens for different JSON fields in the Speakeasy emulated behavioral report training set \cite{quoVadis}. Every additional field included in the analysis increases the vocabulary size. For instance, given filter that uses API calls, file, network, and registry records total vocabulary size is about 2.5M tokens. Given no filters applied, this number jumps close to 8M tokens, exploding vocabulary more than three times and significantly reducing the epistemic density (valuable information per token) of the data.

Concerning field filtering, existing dynamic malware modeling techniques fall into two categories: do not implement any filters \cite{neurlux}, or use only a single type of information, usually API calls \cite{cruparamer, quoVadis}. Focus solely on API calls eliminates valuable behavior representations necessary for establishing effective decision boundaries in malware detection. Research has demonstrated that domain experts rely on a broader range of information when performing actual malware analysis \cite{usenix_inside_re_head}.
Based on ablation studies discussed in \autoref{sec:ablation}, we preserve the following fields from dynamic analysis report: (i) API call names, arguments, and return codes; (ii) file operation type and path; (iii) network connection port and server name; and (iv) registry access type and key value.
We found this combination of fields produces the best generalization and the least overfitting. 




\myparagraph{Normalization.} Retained fields still have unbounded or unpredictable values, which may not inherently contribute to the effectiveness of ML models. For instance, the exact values of IP addresses are not representative \textit{per se} and primarily provide broader context, such as indicating whether the IP is from a private or public network or what autonomous system it belongs to. Similarly, file paths may contain elements like usernames, drive letters, or randomized file and directory names, which have relative contextual significance for behavioral analysis. Hence, the raw values of such fields may not be directly beneficial for modeling, emphasizing the need for suitable normalization before analysis.
We incorporate domain knowledge via placeholders by normalizing filepaths, network connection, and registry access information in the following manner: (i) hash-sums in any field, including SHA1, SHA256, and MD5, are substituted with placeholders like \texttt{<sha1>}, \texttt{<sha256>}, \texttt{<md5>} placeholders; (ii) IP addresses are mapped to placeholders symbolizing loopback, private, public, or IPv6 addresses; (iii) recognizable domain names associated with a list of common top-level domains such as \texttt{com} or \texttt{net} (but not exclusive to these) are assigned the \texttt{<domain>} placeholder; (iv) Windows path variables, for instance, \texttt{\%windir\%} or \texttt{\%userprofile\%}, are expanded to a full path; and (v) frequent Windows paths patterns are replaced with specific placeholders such as \texttt{<drive>} or \texttt{<user>}.

\subsection{Feature Extraction}
We now detail the feature extraction $\vct x = \phi(\vct z^\prime)$ applied after the data cleaning phase $\psi$ by Nebula.

\myparagraph{Tokenization.} 
%
This operation is pivotal for dealing with textual data, since it divide the input text into several atoms named \emph{tokens}, that represent input data in a comprhensible way for machine learning models.
Two basic approaches to tokenization, namely Whitespace and Wordpunct, have been traditionally employed, both using regular expressions to split the text. Former separates words based on spaces, tabs, and newline characters, while latter on top of that uses punctuation as separators. A fragment of whitespace tokenized dynamic analysis report:
\begin{minipage}{\linewidth}
\begin{mdframed}[roundcorner=10pt]
\begin{verbatim}
"0x0", "0x1", "kernel32.getprocaddress",
"0x1000", "0xfa", "kernel32.tlsgetvalue"
\end{verbatim}
\end{mdframed}
\end{minipage}

%

In contemporary deep learning solutions, a more sophisticated approach to tokenization has emerged~\cite{gpt}, predominantly based on Byte Pair Encoding (BPE)~\cite{bpe}, which initially served as a data compression algorithm~\cite{bpeorig}. 
The adoption of BPE as a tokenizer is attributed to its ability to adapt to various languages and tasks seamlessly.
Ideologically, BPE is well suited for machine data such as malware reports, since its data-driven nature allows to learn the optimal tokens scheme directly from the data.
Notably, to handle the intricacies of low-level data in dynamic malware reports, we adjust BPE to incorporate all raw bytes and UTF-8 characters as base tokens. This ensures that event the most rare and unique elements of malware report will have a token-level representation.
The redacted set of BPE tokens covering the same dynamic analysis report fragment are as follows:
\noindent
\begin{minipage}{\linewidth}
\begin{mdframed}[roundcorner=10pt]
\begin{verbatim}
"0x", "0x1", "ne", "32.", "kernel32.", 
"et", "ad", "getproc", "10", "0xf", "tls
\end{verbatim}
\end{mdframed}
\end{minipage}
\\

Furthermore, for both tokenization schemes, we limit our vocabulary to $V=50 000$ most common tokens and introduce two special tokens to denote all other tokens (\texttt{<unk>}) and padding of shorter sequences (\texttt{<pad>}).

\myparagraph{Sequence Length.} In the case of machine data, the tokenized sequences from system log events are typically lengthy. To manage this, we confine behavioral reports to the first N tokens. By keeping the computational budget constant, we evaluate the performance of models with varying sequence lengths. The results of these comparative studies, often referred to as ablation studies, will be detailed in \autoref{sec:ablation}, with the final choice of $N=512$.

\begin{table*}[ht!]
\centering
\caption{Number of samples per malware family in Avast-CTU Dataset\cite{avast_data}.}
\begin{tabular}{l|cccccccccc|c}
    \toprule
    Family & \textbf{Adload} & \textbf{Emotet} & \textbf{HarHar} & \textbf{Lokibot} & \textbf{njRAT} & \textbf{Qakbot} & \textbf{Swisyn} & \textbf{Trickbot} & \textbf{Ursnif} & \textbf{Zeus} & Total \\ \midrule
    Samples & 704 & 14429 & 655 & 4191 & 3372 & 4895 & 12591 & 4202 & 1343 & 2594 & 54000 \\ \bottomrule
\end{tabular}
\label{tab:avast_malware_samples}
\end{table*}

\subsection{Model Architecture}
\label{sec:architecture}

We now detail the last component of Nebula, which is the model function $f$.

\myparagraph{Embedding and Positional Encoding.} Embedding operation maps the input sequence of integers to a higher dimensional space: $e = E(x) \cdot \sqrt{d_e}$, where $E(x)$ is the embedding of the input $x$ and $d_e$ is the dimension of the embedding, with square root used for scaling. This results in vector $e = [e_1, e_2, ..., e_{pos}, ..., e_N ]$, where $e_{pos} \in \mathbb{R}^{d_e}$. 

Since our method relies on the Transformer architecture, which lacks the inherent sense of order provided by recurrent models, we need to incorporate positional information in our sequence. There are multiple alternative ways to encode position. We replicate the approach introduced by Vaswani et al. \cite{transformer_paper}, creating a set of sinusoidal functions with different frequencies for each position in the sequence:
\begin{eqnarray}
PE_{(pos,~2i)} = \sin\left(\frac{pos}{10000^{2i / d}}\right), \\
%
PE_{(pos,~2i+1)} = \cos\left(\frac{pos}{10000^{2i / d}}\right),
\end{eqnarray}
where $PE_{(pos,~i)}$ is the $i$-th dimension of the positional encoding of the token at position $pos$ in the sequence, and $d$ is the dimensionality of the model. The $PE_{(pos,~2i)}$ and $PE_{(pos,~2i+1)}$ terms are used for even and odd dimension $i$ respectively.
These values are then added to the embedded vectors $e_{pos}$ to incorporate the positional information into the sequence
$e_{pos}' = e_{pos} + PE_{pos}$
where $PE_{pos} = [PE_{({pos},~1)}, PE_{({pos},~2)}, ..., PE_{({pos},~d)}]$ is the positional encoding vector for position $pos$. The result is a sequence of vectors $e' = [e_1', e_2', ..., e_N']$, where each vector represents both the token semantics and its position in the sequence, which can now be fed into the Transformer network.

\myparagraph{Neural Layers.} We leverage the Transformer architecture, which originally employs both encoder and decoder layers~\cite{transformer_paper}. Our setup utilizes only encoder layers similar to Devlin et al.~\cite{devlin2019bert}, a design choice that aligns our model with inference task rather than generative objectives as in applications that include decoder \cite{transformer_paper, gpt}. We employ two Transformer encoder layers that align our model size with those of comparable models in \autoref{tab:models}. This choice is not restrictive -- the model can be scaled up to incorporate more Transformer layers to improve performance, consistent with the principle of model scaling laws \cite{scaling_laws}. 
After the self-attention operation, data is forwarded to a classifier for the final prediction. In our implementation, the classifier consists of a fully connected neural network with a single hidden layer composed of $64$ neurons and the final layer for binary or multi-class classification.

\myparagraph{Reduced Self-attention Span.} Input comprised from structured machine data like malware behavior reports contain information in lengthy sequences, which poses a challenge for self-attention architectures like that used by  Transformers~\cite{transformer_paper}, since such models exhibit quadratic complexity with respect to the sequence length. The self-attention operation can be represented as:
\begin{equation}
\label{eq:attention}
\text{Attention}(Q, K, V) = \text{softmax}\left(\frac{QK^T}{\sqrt{d_k}}\right)V,
\end{equation}
where $Q$, $K$, and $V$ as queries, keys, and values, respectively used as inputs to a self-attention layer, and $d_k$ is the dimension of the keys. The product $QK^T$ results in a matrix of size $N \times N$, where $N$ is the sequence length. Calculating this product has a complexity of $O(N^2)$, leading to the quadratic computational complexity with respect to the sequence length.

We propose an alternative approach to reduce computational complexity by partitioning the self-attention operation described in \autoref{eq:attention} into several independent attention spans instead of applying it to the entire sequence. Assume that the original sequence length $N$ is divisible by the span $S$, so there are $M=N/S$ spans. Let $Q_i$, $K_i$, and $V_i$ denote the queries, keys, and values for the $i^{th}$ span. Then, $\text{Attention}(Q_i, K_i, V_i), \quad \forall i \in \{1,2,...,M\}$ and independent attention results are concatenated to vector of size $N$.

In this way, the complexity is reduced to $O(MS^2)$, improving the model's computational efficiency, especially when $S << N$. Our experiments use $S=64$ with $N=512$, resulting in $M=8$ independent self-attention spans. We observe that reducing attention spans enhances the model's inferential capacity on behavioral reports while adhering to the same computational constraints.



\section{Experimental Evaluation}
\label{sec:experiments}
The following section presents an in-depth experimental evaluation designed to assess the effectiveness and robustness of Nebula.
We discuss the dataset used for our experiments (\autoref{sec:data}), and we outline our setup (\autoref{sec:setup}).
We then present our ablation study on the components of Nebula (\autoref{sec:ablation}), followed by its comparison with the state of the art (\autoref{sec:comparison_sota}).
Lastly, we analyse the benefits that self-supervised pre-training has on the required number of data to fit Nebula (\autoref{sec:ssl}), and we conclude by analysing its output with explainability techniques, confirming our findings through domain knowledge (\autoref{sec:explainability}).


\subsection{Datasets}
\label{sec:data}

\begin{table}[t]
    \centering
    \caption{Speakeasy Dataset~\cite{quoVadis} structure and size.}
    \label{table:speakeasy_dataset}
    \begin{tabular}{ccccc}
        \toprule
        & \multicolumn{2}{c}{Training set} & \multicolumn{2}{c}{Test set} \\
        \cmidrule(lr){2-3} \cmidrule(lr){4-5}
        \textbf{Sample label} & \textbf{Size (Gb)} & \textbf{Count} & \textbf{Size (Gb)} & \textbf{Count} \\
        \midrule
        \textit{Benignware} & 127.0 & 26061 & 47.0 & 10000 \\
        Backdoor & 30.0 & 11089 & 7.4 & 2500 \\
        Coinminer & 46.0 & 10044  & 11.0 & 2500\\
        Dropper & 36.0 & 11275  & 9.0 & 2500\\
        Keylogger & 34.0 & 7817  & 9.8 & 2500\\
        Ransomw. & 14.0 & 10014  & 4.6 & 2500\\
        RAT & 5.5 & 9537  & 2.5 & 2500\\
        Trojan & 40.0 & 13128  & 7.1 & 2500\\
        \midrule
        \textbf{Total} & 329 & 98966 & 98 & 27500 \\
        \bottomrule
    \end{tabular}
\end{table}

In our experiments, we evaluate three publicly available datasets by discussing two different types of analysis.

\myparagraph{Malware Detection.}
This binary classification task discerns between benign and malicious software. It's a fundamental task performed by AV and EDR solutions with the aim of detecting malevolent logic running on a system. In real-world applications, it is paramount to maintain severely low false-positive rates to ensure usability and efficiency.

\myparagraph{Malware Classification.}
This is a multi-label classification objective, targeting the attribution of malware samples to a specific type or family. Threat intelligence teams often execute it to study the evolution of malware strains, uncover shared characteristics, and identify potential countermeasures.
We now characterize each dataset according to the best practices established in the malware research~\cite{malware_experiment_practices} by its sample size, the environment used for data collection, its applicability for either malware detection and classification tasks, and the availability of separate training and test sets.

\myparagraph{Speakeasy Dataset~\cite{quoVadis}.} This dataset\footnote{\url{https://www.kaggle.com/ds/3231810}} was generated using Speakeasy v1.5.9~\cite{speakeasy}, a Windows kernel emulator, comprising behavioral reports from in total approximately 93,500 samples, with both legitimate and malicious JSON reports. The malicious samples belong to seven distinct malware types, with sample prevalence across labels detailed in \autoref{table:speakeasy_dataset}. Therefore, the dataset is suitable for both malware detection and classification tasks. The dataset provides a test set explicitly, collected in a different timeframe (April 2022) from the training set (January 2022). This temporal separation facilitates the examination of concept drift in malware behavior.

\myparagraph{Avast-CTU Dataset~\cite{avast_data}.} This dataset\footnote{\url{https://github.com/avast/avast-ctu-cape-dataset}} houses sandbox reports in JSON format derived from CAPEv2~\cite{capev2} (a Cuckoo sandbox~\cite{cuckoo} derivative), with approximately 400,000 samples collected between January 2017 and January 2020. The reports represent ten different malware families (\autoref{tab:avast_malware_samples}). Due to the absence of legitimate samples, this dataset is solely used for malware classification tasks. 
Also, this dataset lacks sequential information, and it only provides a summary of the events colelcted by the sandbox.
The dataset formation aligns with the splitting approach recommended by Bosnansky et al. \cite{avast_data}, in which all samples preceding August 2019 are designated as the training set, while the remainder forms the test set.

\myparagraph{Malicious Code Dataset (MCD)~\cite{kericwy1337_datacon2019_malicious_2019}.} This dataset has approximately 30,000 labeled samples containing API call sequences in XML format without any additional behavioral data (such as filesystem, registry, or network access). The dataset's collection methodology and the environment are not explicitly detailed. The training set contains 10,000 malware and 20,000 goodware samples. As no malware family or type labels are available, this dataset is solely applicable for malware detection task. The test set with 15,000 unlabeled samples cannot be used for evaluation due to the lack of labels. Hence we report mean metrics only on validation sets through cross-validation folds.



\subsection{Experimental Setup}
\label{sec:setup}

Our experiments were conducted on an NVIDIA Quadro T2000, a standard consumer GPU. To align with the limitations of the hardware capacity, the batch size was fixed at $b=96$ for all experiments. For optimization, we employed the AdamW optimizer \cite{adamw} with a static learning rate of $\alpha = 2.5^{-4}$. The hyperparameters were set as $\beta_1 = 0.9$, $\beta_2 = 0.999$, and $\epsilon = 10^{-8}$. An $L_2$ regularization with a weight decay of $\lambda = 1e^{-2}$ was also implemented.
The evaluation metrics were derived from three cross-validation (CV) folds on the training set. The reported metrics are the mean values of the three models evaluated on the validation subsets and a single test set. 
To ensure fair evaluation given the variations in model size as indicated in \autoref{tab:models}, we maintained a constant time budget for training instead of a fixed number of epochs. Each fold was allocated a training duration of five minutes, resulting in a total training budget of 15 minutes per cross-validation run for three folds, excluding pre-processing time. Initial experiments with longer training runs, such as an hour per cross-validation, yielded similar relative outcomes with tolerable deviations.
 As such, the 15-minute training budget was deemed optimal for subsequent experiments.


\subsection{Ablation Studies}
\label{sec:ablation}

We explore here the impact of variations in model components and their configurations on the final performance of the Transformer model. This helps highlight the effectiveness of individual components in the context of the model's overall performance. For our ablation experiments, we use the Speakeasy Dataset as it offers a comprehensive range of behavioral representations. Furthermore, this data enables us to evaluate malware detection performance using a binary classification objective, yielding more interpretable results.

\begin{figure*}[t] 
  \centering
  \begin{subfigure}[b]{0.325\textwidth}
    \includegraphics[width=\textwidth]{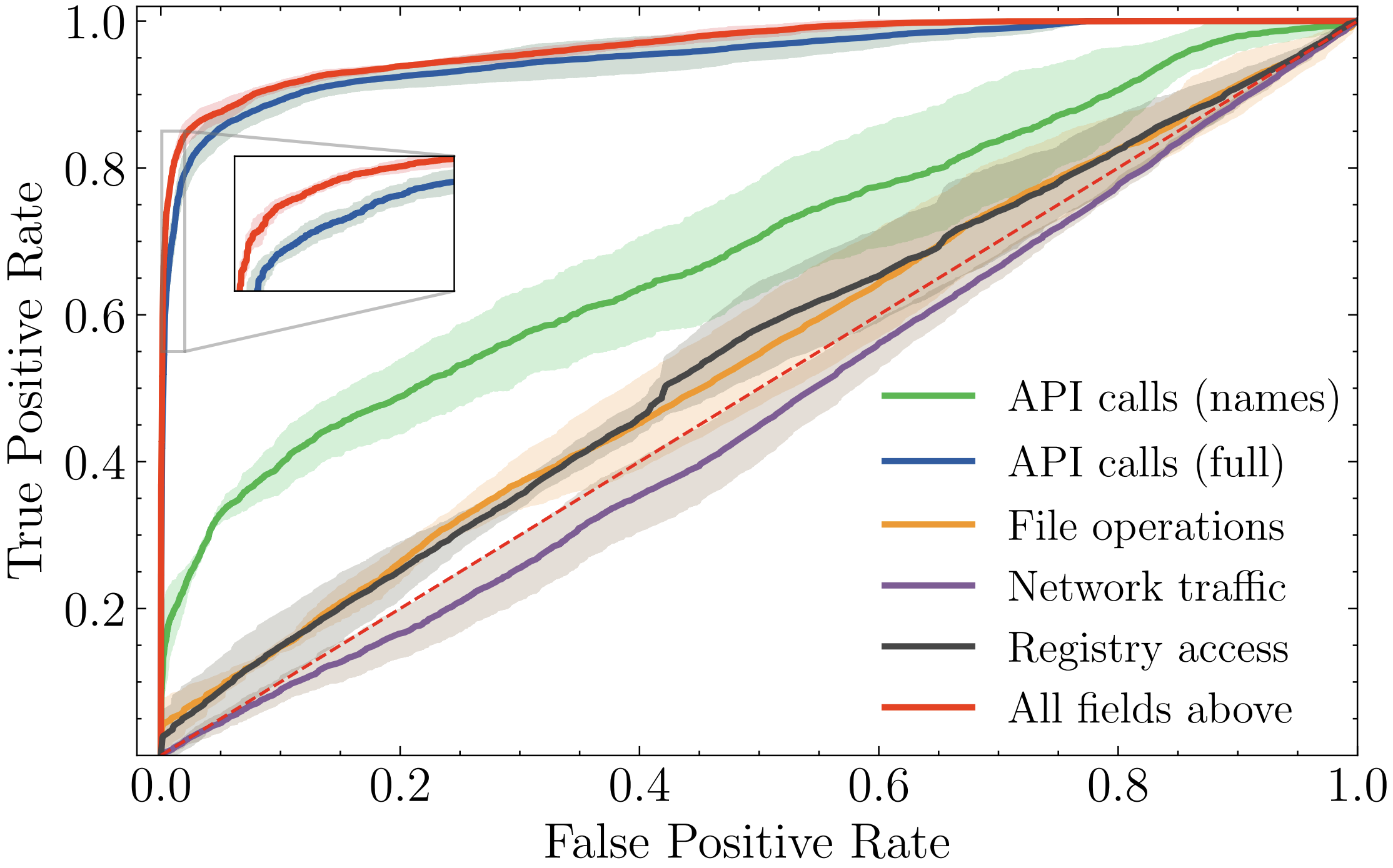}
    \caption{Test set ROC of variable filter fields.}
    \label{fig:field_filter_rocs}
  \end{subfigure}
  \begin{subfigure}[b]{0.325\textwidth}
    \includegraphics[width=\textwidth]{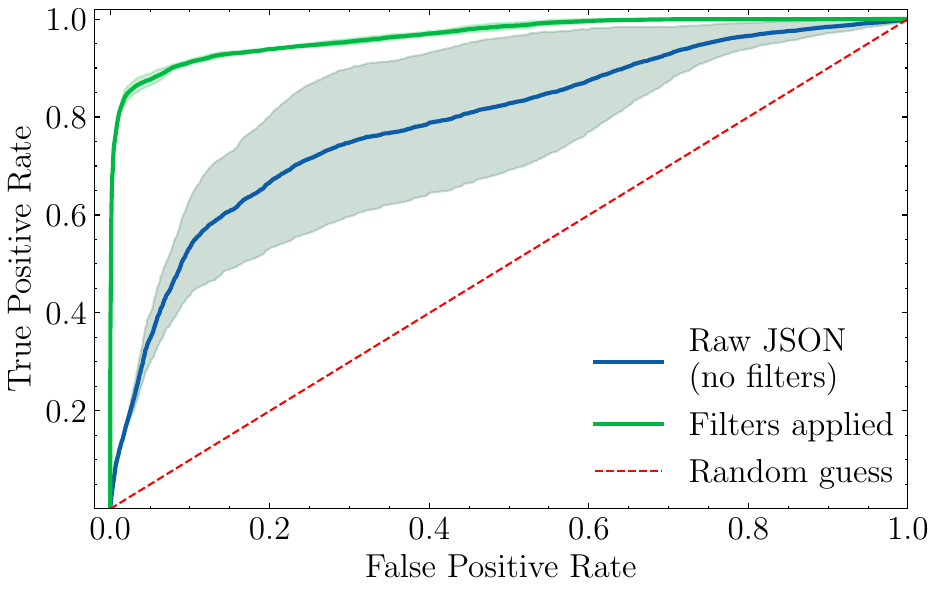}
    \caption{Test set ROC with and without filter setup.}
    \label{fig:filters_no_filters_rocs}
  \end{subfigure}
  \begin{subfigure}[b]{0.325\textwidth}
    \includegraphics[width=\textwidth]{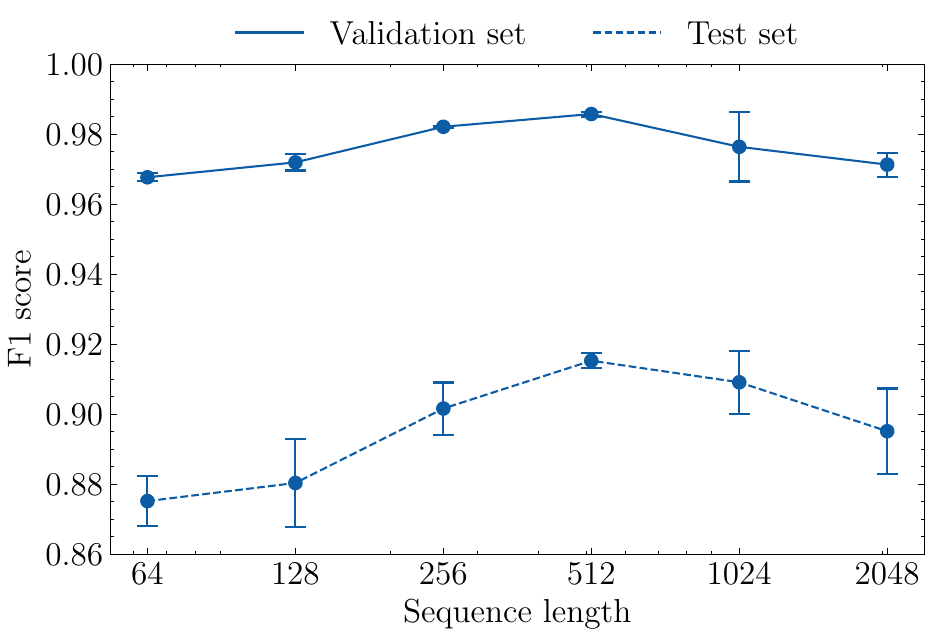}
    \caption{F1 scores with different sequence lengths.}
    \label{fig:seqlens}
  \end{subfigure}
  \caption{Results of ablation studies under different configurations on Speakeasy dataset.}
  \label{fig:ablation_results}
\end{figure*}


\myparagraph{Vocabulary Size.} 
The impact of varying vocabulary size on the performance of the model using the Speakeasy emulation data is presented in \autoref{tab:vocab}. The results demonstrate marginal differences in performance within the range of vocabulary size $V \in \{30~000, ..., 70~000\}$, suggesting that performance in this interval is largely subject to the randomness introduced during model initialization and training. This trend suggests that the model's performance is relatively stable with respect to variations in vocabulary size within this range, indicating a degree of robustness to this parameter. Considering these observations, we chose $V=50~ 000$ as a good compromise that balances performance and complexity.

\begin{table}[t]
    \centering
    \caption{Mean validation set metrics with different vocabulary sizes on malware detection task from Speakeasy data. Reported TPR is at FPR$=10^{-3}$.}
    \begin{tabular}{cccccc}
        \toprule
        Metric & 5k & 10k & 30k & 50K & 70k \\
        \midrule
        TPR & 0.8078 & 0.7834 & \textbf{0.8576} & 0.8383 & 0.8407 \\
        AUC & 0.9965 & 0.9969 & \textbf{0.9977} & 0.9976 & \textbf{0.9977} \\
        F1 & 0.9817 & 0.9839 & 0.9861 & 0.9856 & \textbf{0.9862} \\
        Acc. & 0.9753 & 0.9782 & 0.9811 & 0.9806 & \textbf{0.9814} \\
        \bottomrule
    \end{tabular}
    \label{tab:vocab}
\end{table}

\myparagraph{Field Filters.}
Initially, we examine the utility of individual fields for malware detection. \autoref{fig:field_filter_rocs} presents the outcomes of experiments in which only a specific single field from the behavioral report is retained. Notably, the most influential component of behavioral representation is the sequence of API calls, especially when arguments are provided alongside the API names. All other fields exhibit inferior performance when considered in isolation.
This observation can be rationalized by recognizing that not every type of malware or emulation generates traces in the filesystem, registry, or network -- only a limited subset of emulation reports contain this data. However, all samples invariably exhibit a sequence of API calls, which underscores the critical role of API call information in malware detection.
However, the inclusion of filesystem, registry, or network information in conjunction with API calls enhances detection capabilities. This synergy enables the model to capture a more comprehensive representation of the software's behavior, improving the accuracy and reliability of its predictions.

Additionally, we investigated two preprocessing modalities: (i) a version that abstains from the application of filters, and (ii) one that incorporates optimal field filters during preprocessing. \autoref{tab:filters} presents the F1 scores on the validation and test sets of the Speakeasy emulation reports for both the BPE and whitespace tokenization schemes. 
Remarkably, a significant overfitting issue is present when filters are not employed, evidenced by a difference ($\Delta$) in performance between the validation and test sets. While modeling that employs filters lose about $7\%-8\%$ of F1 on the test set, the performance of modeling without filters degrades down by $23\%-25\%$.
\begin{table}[t]
    \centering
    \caption{Mean F1 values of field filter ablation studies on malware detection task from Speakeasy dataset.}
    \begin{tabular}{lccc}
        \toprule
         Fields & Val. set F1 & Test set F1 & $\Delta$ \\
         \midrule
         Raw JSON (BPE) & 0.9884 & 0.7495 & \textbf{0.2389} \\
         Raw JSON (whtsp.) & 0.9899 & 0.7275 & \textbf{0.2624} \\
         Filtered JSON (BPE) & 0.9847 & 0.9136 & \textit{0.0711} \\
         Filt. JSON (whtsp.) & 0.9870 & 0.9068 & \textit{0.0802} \\
         \bottomrule
    \end{tabular}
    \label{tab:filters}
\end{table}
A visual examination of this trend is depicted in \autoref{fig:filters_no_filters_rocs}, where the Receiver Operating Characteristic (ROC) curve for the test set demonstrates significant degradation when filters are not employed. Additionally, the high standard deviation between cross-validation runs suggests a level of model instability or variance in prediction.
The observed outcome can be attributed to the presence of unconstrained variables representative of one specific execution, like hash sum or start address memory segment. These fields cause the model to overfit the training data, hindering its generalization and predictive capabilities to unseen data in the test set. Hence, the application of field filters appears instrumental in enhancing model stability and performance, contributing to more reliable and generalizable predictions.
\begin{table}[t]
    \centering
    \caption{Mean test set metrics of tokenizer ablation studies on malware detection task from Speakeasy data. Reported TPR is at FPR$=10^{-3}$.}
    \begin{tabular}{ccccc}
        \toprule
        Tokenizer & TPR & AUC & F1 & Acc.  \\
        \midrule
        Wordpunct & 0.5540 & 0.9630 & 0.9049 & 0.9041 \\ 
        Whitespace & \textbf{0.5703} & \textbf{0.9664} & 0.9068 & 0.9053 \\
        BPE & 0.5213 & 0.9657 & \textbf{0.9136} & \textbf{0.9104} \\
        \bottomrule
    \end{tabular}
    \label{tab:tokenizer_results}
\end{table}

\myparagraph{Tokenization.} We conducted ablation studies on tokenization to investigate the impact of different tokenization strategies on model performance. Three different tokenizers were tested: BPE~\cite{sentencepiece}, Whitespace, and Wordpunct~\cite{nltk_lib}. The test set F1 scores for different tokenization methods are reported in \autoref{tab:tokenizer_results}.
The results reveal that all three tokenization methods deliver comparable mean F1 scores. 
The BPE tokenizer demonstrates slightly better generalization capabilities, achieving an F1 score on the test set that is almot 1\% higher than the others. 
This observation is further supported by the field filter experiments discussed in the previous paragraph, with results in \autoref{tab:filters}, where BPE exhibited the smallest performance decrease ($\Delta$) between the validation and test sets.
Furthermore, it is noteworthy that the Whitespace tokenizer achieves impressive results on the test set, surpassing the other tokenization methods if evaluated by area under the curve (AUC) or true positive rate (TPR) at false positive rate (FPR) of $10^{-3}$, as shown in \autoref{tab:tokenizer_results}. Given competitive performance of BPE and Whitespace, we report metrics of both tokenizers for subsequent malware detection and classification, as well as explainability experiments.

\begin{table}[t!]
    \centering
    \caption{Malware detection metrics on Speakeasy test dataset. Reported TPR is at FPR$=10^{-3}$.}
    \resizebox{\linewidth}{!}{%
    \begin{tabular}{l c c c c c}
        \toprule
        & & \multicolumn{4}{c}{Test set} \\
        Model & Training batches & TPR & AUC & F1 & Acc. \\
        \midrule
        Gated CNN~\cite{zhang_dmds} & \underline{1058} & 0.2152 & 0.8879 & 0.6465 & 0.7014 \\
        Neurlux~\cite{neurlux} & 7406 & 0.4250 & 0.9528 & 0.8792 & 0.8786 \\
        Quo.Vadis~\cite{quoVadis} & 4761 & 0.3081 & 0.9224 & 0.8065 & 0.8173 \\
        \midrule
        Nebula (BPE) & \textbf{2116} & 0.5213 & 0.9657 & \textbf{0.9136} & \textbf{0.9104} \\
        Nebula (whitesp.) & 2159 & \textbf{0.5703} & \textbf{0.9664} & 0.9058 & 09053 \\
        \bottomrule
    \end{tabular}}
    \label{tab:results_speakeasy}
\end{table}
%
\begin{table}[t!]
    \centering
    \caption{Malware detection metrics on MCD test dataset. Reported TPR is at FPR$=10^{-3}$.}
    \begin{tabular}{lcccc}
    \toprule
         Model & TPR & AUC & F1 & Acc.  \\
         \midrule
         Neurlux~\cite{neurlux} & 0.8508 & 0.9942 & \textbf{0.9687} & \textbf{0.9794} \\
         Quo.Vadis~\cite{quoVadis} & \textbf{0.9035} & \textbf{0.9950} & 0.9613 & 0.9736 \\
         \midrule
         Nebula (BPE) & 0.8332 & 0.9937 & 0.9653 & 0.9770 \\
         Nebula (whitesp.) & 0.8243 & 0.9932 & 0.9590 & 0.9731 \\
         \bottomrule
    \end{tabular}
    \label{tab:cruparamer_results}
\end{table}

\myparagraph{Sequence Length.} \autoref{fig:seqlens} depicts the F1 scores on the validation and test sets with varying sequence lengths. The performance on both validation and test sets peaks at a sequence length of $N=512$. This suggests that sequences of length $N \in \{64, ..., 256\}$ may not encapsulate all the necessary information for effective model inference, leading to a significant drop in test set performance. On the other hand, longer sequences are more computationally demanding, especially when utilizing self-attention-based modeling. Hence, under the same computational time constraints, sequences with length $N \in [1024, 2048]$ yield less robust results.

\subsection{Comparison with State of the Art}
\label{sec:comparison_sota}

\begin{figure}[t]
    \centering
    \includegraphics[width=0.95\linewidth]{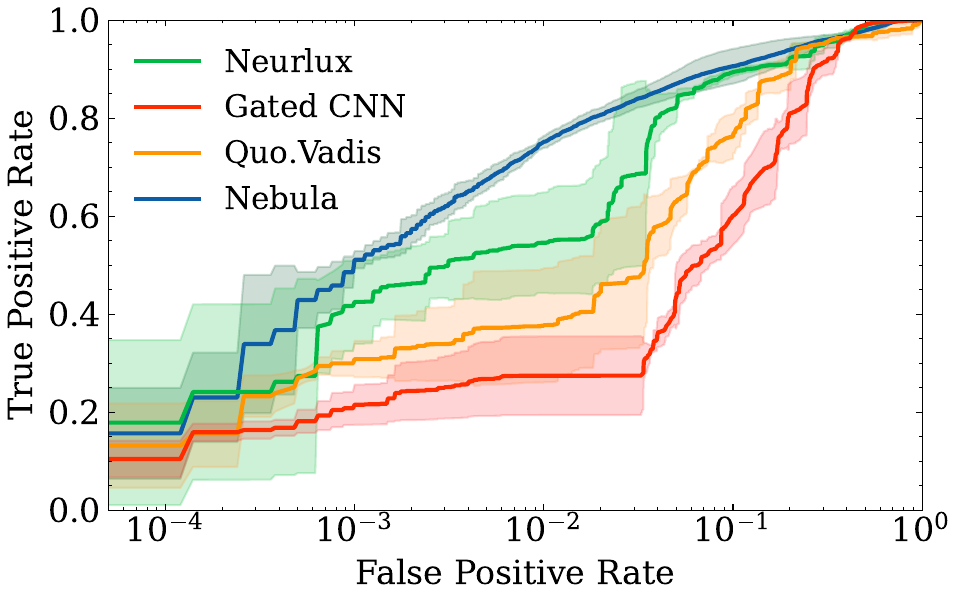}
    \caption{Mean ROC curves over three cross-validations for malware detection of Neurlux~\cite{neurlux}, Gated CNN~\cite{zhang_dmds}, Quo.Vadis~\cite{quoVadis} and Nebula (with BPE tokenizer) on Speakeasy data test set.}
    \label{fig:results_roc_speakeasy_test}
\end{figure}

\myparagraph{Malware Detection.} 
In this section, we evaluate the performance of Nebula, with alternative models in the domain of malware detection. Metrics on the Speakeasy dataset~\cite{quoVadis} reported in \autoref{tab:results_speakeasy}. ROC curve on the test set exemplified in \autoref{fig:results_roc_speakeasy_test}. Four modeling techniques were able to model this data, namely Neurlux presented by Jindal et al.~\cite{neurlux}, Gated CNN model by Zhang et al.~\cite{zhang_dmds}, Quo.Vadis released by Trizna~\cite{quoVadis}, and Nebula.
Our model surpasses all competitive architectures on Speakeasy emulation data, outperforming all metrics on the test sets in either the whitespace and BPE tokenization modes. This is particularly evident under low false-positive conditions. For instance, with $ 10^{-3}$ FPR, Nebula with whitespace tokenization demonstrates $0.570$ TPR on the test set.
In comparison, the next best performing model, Neurlux, scores $0.42$ TPR on the test set.
This observation becomes critically significant considering that strict low false positive rates are enforced on production-grade malware detectors \cite{edr}. 

\begin{table*}[t]
\centering
\caption{Mean F1 scores for malware classification objective on Spekeasy dataset.}
\begin{tabular}{lcccccccc}
\toprule
 & Clean & Backdoor & Coinminer & Dropper & Keylogger & Ransomware & RAT & Trojan \\
\midrule
Neurlux & 0.8453 & 0.8329 & \textbf{0.6910} & 0.4488 & 0.2032 & 0.5527 & \textbf{0.6625} & 0.6153 \\
Gated CNN & 0.7588 & 0.6870 & 0.5586 & 0.2015 & 0.0794 & 0.3584 & 0.0000 & 0.5282 \\
Quo.Vadis & 0.8338 & 0.8520 & 0.4884 & 0.3580 & \textbf{0.2119} & 0.6861 & 0.1195 & 0.5359 \\\midrule
Nebula (BPE) & \textbf{0.8526} & \textbf{0.8548} & 0.6303 & 0.2850 & 0.1295 & \textbf{0.7421} & 0.3683 & \textbf{0.6827} \\
Nebula (whitsp.) & 0.8240 & 0.8324 & 0.6214 & \textbf{0.4615} & 0.1179 & 0.6523 & 0.1854 & 0.6486 \\
\bottomrule
\end{tabular}
\label{tab:f1_scores}
\end{table*}
\begin{table*}[t]
\centering
\caption{Mean F1 scores for malware classification objective on Avast-CTU dataset.}
\begin{tabular}{lcccccccccc}
\toprule
 & Adload & Emotet & HarHar & Lokibot & Qakbot & Swisyn & Trickbot & Ursnif & Zeus & njRAT \\
\midrule
Neurlux & \textbf{0.7150} & 0.9294 & \textbf{0.9031} & 0.8320 & 0.9320 & \textbf{0.9991} & \textbf{0.9536} & 0.8910 & 0.6503 & 0.8479\\ \midrule
Nebula (BPE) & 0.4390 & \textbf{0.9392} & 0.7763 & 0.8957 & \textbf{0.9876} & 0.9973 & 0.9227 & 0.9362 & 0.6419 & 0.8656 \\
Nebula (whitsp.) & 0.6975 & 0.9319 & 0.8363 & \textbf{0.9048} & 0.9768 & 0.9984 & 0.9056 & \textbf{0.9585} & \textbf{0.6690} & \textbf{0.8896} \\
\bottomrule
\end{tabular}
\label{tab:f1_scores_family}
\end{table*}
The efficiency of Nebula is also reflected in the number of training batches required. As seen in \autoref{tab:results_speakeasy}, Nebula achieves these results with less than a third of the training batches required by the second-best model, Neurlux.
Turning awareness to the Malware Code Dataset (MCD)~\cite{kericwy1337_datacon2019_malicious_2019}, the mean validation set metrics are presented in \autoref{tab:cruparamer_results}. MCD preprocessing is computationally demanding due to the high information density per sample because of lengthy API call traces. This has a detrimental effect on models that employ custom feature engineering schemes, such as Zhang et al. \cite{zhang_dmds}, which take several seconds of feature engineering per MCD sample. Processing a training set of 30,000 samples in this manner would take approximately 100 hours—impractical in both experimental and real-world scenarios. Consequently, we excluded this model from our experiments on MCD.

Our observations reveal that Quo.Vadis, a simplistic modeling scheme focused solely on API call names, outperforms both Neurlux and Nebula based on AUC and detection rate under low false-positive conditions. Given the lengthier API call sequences in the MCD data, narrowly focused models like Quo.Vadis might capture more behavior relevant to the data. This outlines the evidence that narrow modeling schemes are still more tuned to this specific data type for specific data sources and can outcompete more general mechanisms.

\myparagraph{Malware Classification.} Predicting malware family is a multi-label objective, and we report the results of the performances of the considered models in \autoref{tab:f1_scores} and \autoref{tab:f1_scores_family} the F1 scores on the Speakeasy and Avast-CTU datasets. 
%
%
Due to the lacking of sequential information of the Avast-CTU dataset, we omit Quo.Vadis and Gated CNN from the comparison, as they require temporal information.
Thus, we only evaluate this dataset with Neurlux and Nebula.
Thus Avast-CTU analysis includes these models only.

\begin{figure}
    \centering
    \includegraphics[width=0.95\linewidth]{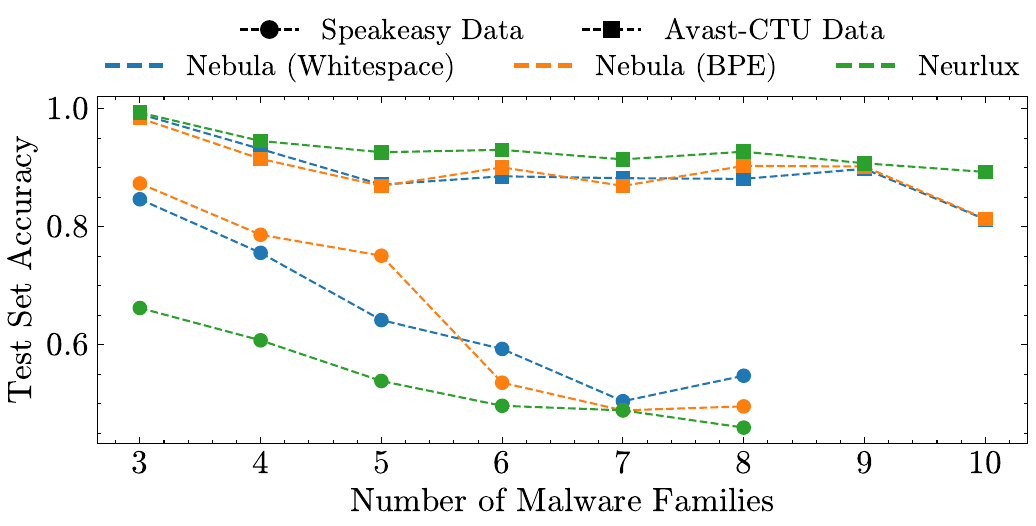}
    \caption{Test set accuracies with variable number of families used for malware classification task.}
    \label{fig:family_downscale}
\end{figure}

Nebula exhibits superior test set F1 scores for 4 out of 7 malware types on Speakeasy (\autoref{tab:f1_scores}) data and in 6 out of 10 malware families on Avast-CTU data (\autoref{tab:f1_scores_family}).
This is particularly noticeable in malware families experiencing significant concept drift, such as polymorphic Emotet~\cite{cisa2020emotet}, in families with many sub-variants, like Zeus~\cite{krebs2014zeus}, or on malware types that exhibit rich and diverse behaviors, such as benignware, backdoors, ransomware, or trojans. \textit{Modus operandi} of such agents require frequent manipulation with network, filesystem, and registry.
An examination of metrics on Speakeasy Dataset test set shows that Neurlux still surpasses Nebula in detecting Droppers and RATs, achieving $18\%$ and $30\%$ higher F1 scores, respectively. 
This might suggest a weakness in Nebula's data-cleaning approach for these particular malware families, indicating a potential avenue for future improvements. 
Simultaneously, models focusing solely on API calls, for example, Quo.Vadis, exhibit slightly superior performance over the general models for malware families with less diverse behavior, such as in detection of Keyloggers, a malware type that only occasionally interacts with the network or filesystem to store logged keys.
Also, we analyze the performances under varying number of target families in malware classification task, depicting the test set accuracies in \autoref{fig:family_downscale}.
Performance of malware classification capabilities drops as number of families grow, suggesting that in practical threat intelligence (TI) applications, it is supposedly better to employ numerous models, each tailored to identify specific key malware families and classes, instead of relying on general classifiers.
While the variability of performance for the Avast-CTU dataset remains relatively minor, with only few percentage points of difference, the performance variance on Speakeasy reports notably diverges.
Nebula demonstrates notably superior performance to Neurlux, particularly in scenarios involving a smaller number of families, exhibiting at least a 20\% accuracy advantage in Nebula's BPE model over Neurlux in tasks encompassing 3-5 target families.
This may prove to be particularly valuable in practice, reinforcing the observation for tailored models targeting a lower number of families for optimal performance in TI tasks.

\subsection{Self-Supervised Learning Benefits}
\label{sec:ssl}

Since Nebula leverages transformers, we now exploreits capacity for self-supervised learning (SSL), by leveragin unlabeled data to pre-train models. 
This is achieved through language modeling (LM), with two prominent strategies prevailing in textual data processing: masked language modeling, as exemplified on BERT and related transformer-encoder architectures~\cite{devlin2019bert}, and autoregressive next-token prediction, characteristic of generative tasks like GPT models~\cite{gpt}.
For our study in malware detection, we conduct experiments on both techniques, and we evaluate the performance compared to the fully-supervised settings.
Since LM tasks produce logits in size of vocabulary for these experiments we decreased vocabulary size to 8192 for computational reasons. 
As autoregressive LM requires global attention, we employed a Transformer architecture tailored for these experiments, which discards attention chunking as discussed in \autoref{sec:architecture}. As for masked LM, we employed the same pre-training parameters as in BERT~\cite{devlin2019bert} setup.
We designated 80\% of the training data as an unlabeled corpus for self-supervised pre-training, while the remaining 20\% was allocated for supervised fine-tuning. To provide context, we included two benchmarks as proposed by Apruzzese et al.~\cite{sok_apruzzese}: (i) an upper bound, representing a supervised model trained on the full dataset with access to all label information, and (ii) a lower bound supervised model that undergoes no pre-training and utilizes only 20\% of the training set, akin to the fraction used for LM fine-tuning.

\begin{figure}
    \centering
    \includegraphics[width=\linewidth]{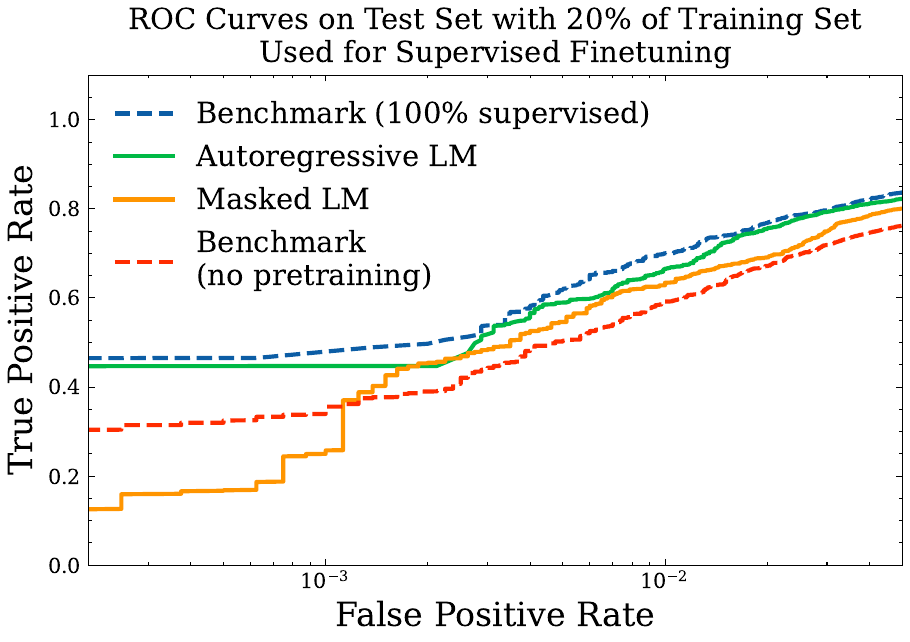}
    \caption{Self-supervised language modeling (LM) efficiency with 80\% of training set used for pre-training and 20\% for supervised fine-tuning, compared with two benchmarks without self-supervised LM, representing upper bound with access to all the labels (100\% supervised) and lower bound using only fine-tuning data (20\% supervised).}
    \label{fig:pretraining}
\end{figure}

We report ROC curves on test set for all runs in \autoref{fig:pretraining}, and we observe a consistent performance pattern across models. As anticipated, the upper bound model exhibits the highest detection rates, while the lower bound model performs the least effectively, with ~15\% gap between both model detection rates, indicative of the significance of the additional 80\% of training data available to the upper bound model. The masked LM model demonstrates the second-worst performance, particularly under the strictest conditions of FPR$=10^{-3}$, reporting detection rates even inferior to those of the lower bound model. This discrepancy suggests that masked LM pre-training may learn detrimental representations that remain insufficiently adjusted during fine-tuning. In contrast, the autoregressive LM model yields remarkable results, nearly matching the performance of the upper bound supervised model across all FPR ranges and particularly closely aligning with it under the lowest FPR, with only a minimal 3\% drop in detection rate. This finding suggests that Nebula can effectively leverage substantially less labeled data by consuming unlabeled samples, thereby reducing human resource requirements and enabling the utilization of vast amounts of PE and DLL files available to community and private businesses in the process.

\subsection{Explaining the Behavior of Nebula}
\label{sec:explainability}

\begin{table*}[ht!]
\centering
    \caption{Top 10 individual tokens impacting the decision of Nebula towards the target class according to integrated gradients~\cite{sundararajan2017axiomatic} XAI methodology based on 1000 samples per class from Speakeasy dataset. Normalized importance scores for each token are reported in parentheses.}
    \input{table2}
    \label{tab:explanation}
\end{table*}


We now explain the behavior of Nebula leveraging two \emph{explainable AI} (XAI) techniques. 
The first one is \emph{Integrated Gradients}~\cite{sundararajan2017axiomatic}, that computes the importance of input features by integrating gradients along a path from a baseline to the input. 
In our case, we use an empty JSON file as the baseline, which stands for the absence of any behavior.
We leverage the GradientSHAP implementation from the SHapley Additive exPlanations (SHAP) framework~\cite{shap_library}.
Since this technique requires an end-to-end differentiable model, it is not directly applicable in our case due to the presence of the initial embedding layer. To overcome this issue, we extract sample embeddings and obtain explanations from this point onwards, recovering importance values by taking the mean over the embedding dimension. 
As the second method, we leverage attention activations from the transformer encoder layer to indicate the learned importance of relative token weights within the model. Transformer self-attention layers are multi-headed; in our case, each layer has eight independent heads. We examine all the layers and heads, focusing on the strongest attention weight deviations and investigating their implications.
We perform a large-scale analysis, randomly subsampling 1000 samples for each malware type from the Speakeasy dataset. The results are shown in \autoref{tab:explanation}, from which we can derive the following key aspects:

\begin{itemize}
    \item Functions from the \textit{advapi32} DLL exhibit a pronounced significance in malware detection. This library provides functionalities that allow programs to interact with the OS, for instance, seeking elevated privileges or manipulating service controls. The prominence of these functions in our findings underscores their recurrent misuse in malware.
    \item The token \verb|0xcf0000| is used solely with \texttt{CreateWindowEx} API call, referencing the style of spawned window: \texttt{WS\_OVERLAPPEDWINDOW}. We see that the general trend for malware samples in the test set involves the initiation of user interaction with this specific parameter of UI behavior.
    \item API calls like \texttt{setenvnvar} (alias for \textit{SetEnvironmentVar}) and \texttt{getsockobj} (alias of \textit{GetSocketObject}) are frequently among the top tokens for malware classification.
    This indicates the necessity of malware for frequent manipulation with environment variables, and the need to make network connections, as well as an indication of these manipulations as valuable components for the final heuristic by Nebula.
\end{itemize}

\begin{figure}[t]
    \centering
    \begin{subfigure}[b]{\linewidth}
        \includegraphics[width=0.95\linewidth]{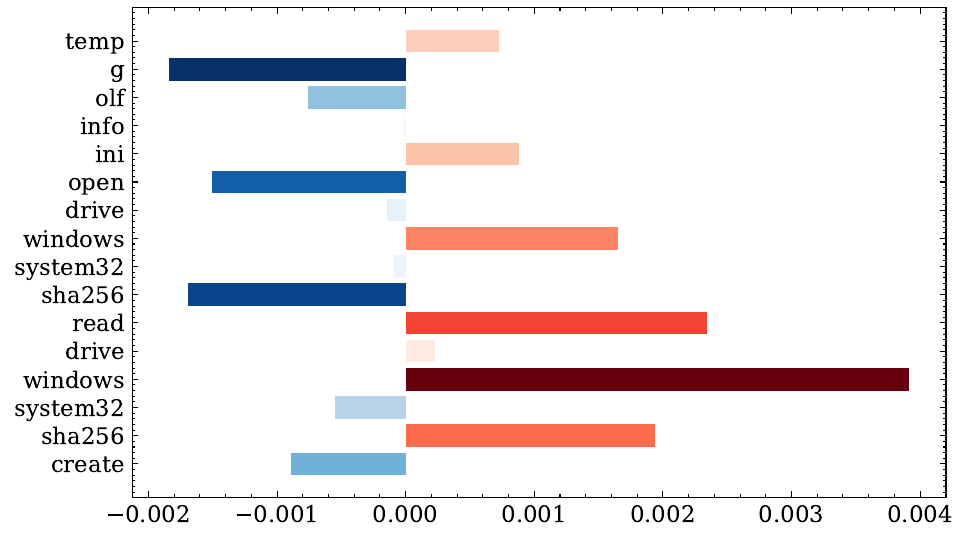}
        \caption{Explainability analysis based on integrated gradients~\cite{sundararajan2017axiomatic} method.}
        \label{fig:backdoor_shap}
    \end{subfigure}
    \begin{subfigure}[b]{\linewidth}
        \centering
        \includegraphics[width=0.5\linewidth]{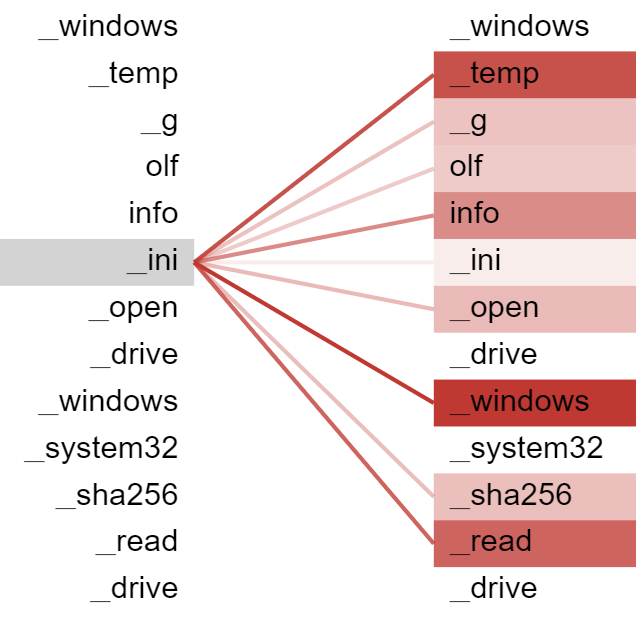}
        \caption{Attention activations~\cite{bertviz} at the second self-attention layer.}
        \label{fig:backdoor_attention}
    \end{subfigure}
    \caption{Depiction of fragment from ``Urelas'' trojan dynamic analysis report exhibiting filesystem interactions. Both explainability technique agree on the importance (red) of tokens like \texttt{windows, temp, read}, all linked to filesystem activities likely exploited by the analysed malware.}
    \label{fig:shap_attention}
\end{figure}

Notably, while \autoref{tab:explanation} reports the token importance in isolation, these are used by Nebula's self-attention mechanism in relation to all other tokens in sequence. We manually ensure that the importance assigned by the integrated gradients method from SHAP library~\cite{sundararajan2017axiomatic} is directly correlated with attention weights from within Transformer encoder heads. For instance, we showcase the results of both XAI techniques on a specific sample infected with the ``Urelas'' trojan (SHA1: \textit{c7ee95f0ea78400d5e4938e06fea1bb0c388b565}) in \autoref{fig:shap_attention}. 
We find that both integrated gradients and attention activations identify the highest maliciousness indicators within a particular dynamic analysis segment shown in \autoref{fig:shap_attention}, pinpointing the same tokens representing filesystem manipulations as highly associated with maliciousness. 

\section{Related Work}
\label{sec:related}

We are not the first to explore Transformer applicability for malware detection, by also discussing the applicability of self-attention only for \textit{static} malware analysis, contrary to our contribution on dynamic malware analysis.
Li et al.~\cite{imad} were the first to propose a Transformer-based architecture for static malware analysis applied on assembly instructions. They used a custom architecture called ``Galaxy Transformer'' to avoid length limitations and construct hierarchical representations.
Rudd et al. \cite{mandiant_transformer} explored Transformer applicability on static malware detection applied on raw malware bytes. Influenced by the success of the GPT modeling scheme~\cite{gpt}, the authors analyzed Transformer decoder with an autoregressive pre-training objective.
Pei et al. \cite{pei_similarity_dynamic_traces} apply a hierarchical Transformer for code similarity analysis and vulnerability detection. They generate a dataset from benign Linux ELF binaries, obtaining behavioral micro-traces with QEMU based Unicorn emulator.
Similarly to our work, existing approaches~\cite{r3_reference1_ensemble, r3_reference2} explored the usage of transformers applied on sequence of API calls, and comparing them with alternative architectures.
However, the application of Transformers customized to a dynamic malware context and applicable to variety of telemetries, distinguishes our approach. Moreover, the comparison with other SotA dynamic malware detectors and our exploration of the model's explainability represent additional, distinct contributions with respect to previous work.

\section{Conclusions, Limitations, and Future Work}
\label{sec:conclusions}
In this paper, we present Nebula, a novel self-supervised learning transformer model for dynamic malware detection and classification, and we select its components through an extensive ablation study. 
We show how much the inclusion of different behavioral aspects manifested by malware improves the performance, by also quantifying how much data cleaning procedure boosts the accuracy at test time.
We compare our approach against previously-proposed machine-learning methods for dynamic malware analysis, in a pure supervised learning setting, and we show that Nebula often achieves better results than CNNs and LSTMs.
In particular, Nebula surpasses, on average, its competitors on both malware detection and classification on three different datasets.
We study how self-supervised pre-training can reduce the need for training data, highlighting that the usage of only 20\% of the training dataset is enough to reach state-of-the-art performance on malware detection.
Lastly, we inspect the output of Nebula through two explainability methods, and we reveal that Nebula is giving attention to relevant tokens associated with malicious activity, by also exhibiting long spans of attention.

\myparagraph{Limitations.}
We have not considered robustness of our model against adversarial malware~\cite{demetrio2021adversarial}. We acknowledge that such analysis would be of high interest, however, adversarial attack on dynamic classifier in input space would require algorithmic modification of malware sample without corrupting the malicious logic. To date, only initial methods of such perturbations have been explored~\cite{rosenberg2018generic}, with no released implementation to replicate these attacks.
Studying and implementing attacks against behavioral classifiers would stand as a contribution on its own, and for this reason we only discuss them as a constraint of our study.
Another limitation of Nebula is reliance on quality of dynamic analysis. Some malicious samples will refrain expressing malicious logic given the execution in virtualized or emulated environments. Techniques focused on sandbox evasion techniques~\cite{sandbox_evasion} will reduce quality of Nebula even more, emphasizing the need of hybrid heuristic that incorporates signature, static, and dynamic methods given deployment in production setting~\cite{quoVadis}.
Lastly, we caution that pre-trained Nebula models we release were trained on just 70k samples and not a general pre-trained malware detectors with real-world predictive power, instead, valuable for further research and experiments on that same dataset only.

\myparagraph{Future work.}
We plan to further investigate the effect of self-supervised learning on Nebula, pre-training it on a much larger data collection with unlabeled samples, and by varying the size of labelled data.
We hence envision models trained with a scarcely populated dataset of novel malware, speeding up computations and keeping state-of-the-art performance that permit the deployment of these novel technologies.
Finally, we emphasize importance of assessing the adversarial robustness properties of Nebula, seeing high potential in future work on developing novel attack algorithms tailored to bypass dynamic malware detectors and classifiers.

\bibliographystyle{unsrt}
\bibliography{library}

\appendix

\section{Malware Classification Result Tables}
\label{apndx:tables}

Below we present detailed malware classification results with evaluation metrics per each malware type on Speakeasy Dataset~\cite{quoVadis} in \autoref{tab:malware_classificatio_speakeasy} and per each malware family on Avast-CTU Dataset~\cite{avast_data} in \autoref{tab:malware_classification_avast}.

\begin{table*}[!htb]
    \caption{Malware classification results on Speakeasy Dataset. Reported TPR is at FPR$=10^{-3}$.}
    \label{tab:malware_classificatio_speakeasy}
    \begin{subtable}{.5\linewidth}
      \centering
        \caption{Validation set}
        \begin{tabular}{lcccc}
            \toprule
            Model & TPR & AUC & F1 & Accuracy \\
            \midrule
		\multicolumn{5}{c}{Benignware} \\
		\midrule
		Gated CNN~\cite{zhang_dmds} & 0.114 & 0.9209 & 0.8793 & 0.9166 \\
		Neurlux~\cite{neurlux} & 0.3486 & \textbf{0.9756} & 0.9609 & 0.9741 \\
		Quo.Vadis~\cite{quoVadis} & \textbf{0.4605} & 0.958 & 0.9463 & 0.9655 \\
		Nebula (BPE) & 0.4475 & 0.9729 & \textbf{0.9613} & \textbf{0.9748} \\
		\midrule
		\multicolumn{5}{c}{Backdoor} \\
		\midrule
		Gated CNN~\cite{zhang_dmds} & 0.6974 & 0.8456 & 0.8059 & 0.952 \\
		Neurlux~\cite{neurlux} & 0.6586 & \textbf{0.9223} & 0.8929 & 0.9704 \\
		Quo.Vadis~\cite{quoVadis} & \textbf{0.7898} & 0.8924 & 0.8711 & 0.9661 \\
		Nebula (BPE) & 0.775 & 0.9209 & \textbf{0.8965} & \textbf{0.9716} \\
		\midrule
		\multicolumn{5}{c}{Coinminer} \\
		\midrule
		Gated CNN~\cite{zhang_dmds} & 0.553 & 0.8917 & 0.801 & 0.9634 \\
		Neurlux~\cite{neurlux} & \textbf{0.8938} & \textbf{0.9445} & \textbf{0.923} & \textbf{0.9865} \\
		Quo.Vadis~\cite{quoVadis} & 0.8477 & 0.9288 & 0.8811 & 0.9788 \\
		Nebula (BPE) & 0.8804 & 0.9371 & 0.9083 & 0.9839 \\
		\midrule
		\multicolumn{5}{c}{Dropper} \\
		\midrule
		Gated CNN~\cite{zhang_dmds} & 0.2407 & 0.8793 & 0.7665 & 0.9477 \\
		Neurlux~\cite{neurlux} & \textbf{0.4505} & \textbf{0.9659} & \textbf{0.8895} & \textbf{0.9741} \\
		Quo.Vadis~\cite{quoVadis} & 0.3434 & 0.959 & 0.8704 & 0.9695 \\
		Nebula (BPE) & 0.4148 & 0.9482 & 0.8742 & 0.9714 \\
		\midrule
		\multicolumn{5}{c}{Keylogger} \\
		\midrule
		Gated CNN~\cite{zhang_dmds} & 0.4601 & 0.8403 & 0.7177 & 0.9689 \\
		Neurlux~\cite{neurlux} & \textbf{0.6925} & 0.9503 & \textbf{0.8576} & \textbf{0.9826} \\
		Quo.Vadis~\cite{quoVadis} & 0.3644 & \textbf{0.9577} & 0.7961 & 0.9723 \\
		Nebula (BPE) & 0.5685 & 0.9376 & 0.8293 & 0.9789 \\
		\midrule
		\multicolumn{5}{c}{Ransomware} \\
		\midrule
		Gated CNN~\cite{zhang_dmds} & 0.9252 & 0.9587 & 0.9356 & 0.9839 \\
		Neurlux~\cite{neurlux} & 0.961 & 0.9793 & 0.9732 & 0.9933 \\
		Quo.Vadis~\cite{quoVadis} & \textbf{0.9839} & \textbf{0.9908} & \textbf{0.9844} & \textbf{0.9961} \\
		Nebula (BPE) & 0.9713 & 0.9833 & 0.9699 & 0.9924 \\
		\midrule
		\multicolumn{5}{c}{RAT} \\
		\midrule
		Gated CNN~\cite{zhang_dmds} & 0.0381 & 0.5142 & 0.052 & 0.9782 \\
		Neurlux~\cite{neurlux} & \textbf{0.5047} & \textbf{0.7497} & \textbf{0.6564} & \textbf{0.9884} \\
		Quo.Vadis~\cite{quoVadis} & 0.2861 & 0.6387 & 0.3688 & 0.9806 \\
		Nebula (BPE) & 0.4171 & 0.7053 & 0.5618 & 0.9857 \\
		\midrule
		\multicolumn{5}{c}{Trojan} \\
		\midrule
		Gated CNN~\cite{zhang_dmds} & 0.1253 & 0.8425 & 0.6891 & 0.9156 \\
		Neurlux~\cite{neurlux} & \textbf{0.3903} & 0.8827 & \textbf{0.8022} & \textbf{0.9516} \\
		Quo.Vadis~\cite{quoVadis} & 0.2007 & 0.8788 & 0.7666 & 0.9391 \\
		Nebula (BPE) & 0.2268 & \textbf{0.8878} & 0.7853 & 0.9442 \\
            \bottomrule
        \end{tabular}
    \end{subtable}%
    \begin{subtable}{.5\linewidth}
      \centering
        \caption{Test Set}
        \begin{tabular}{lcccc}
            \toprule
            {} & TPR & AUC & F1 & Accuracy \\
            \midrule
		\multicolumn{5}{c}{Benignware} \\
		\midrule
		Gated CNN~\cite{zhang_dmds} & 0.0229 & 0.7482 & 0.7588 & 0.736 \\
		Neurlux~\cite{neurlux} & 0.0347 & 0.8464 & 0.8453 & 0.8362 \\
		Quo.Vadis~\cite{quoVadis} & \textbf{0.0395} & 0.8385 & 0.8337 & 0.8327 \\
		Nebula (BPE) & 0.0382 & \textbf{0.8555} & \textbf{0.8526} & \textbf{0.8468} \\
		\midrule
		\multicolumn{5}{c}{Backdoor} \\
		\midrule
		Gated CNN~\cite{zhang_dmds} & 0.4102 & 0.7894 & 0.687 & 0.9395 \\
		Neurlux~\cite{neurlux} & 0.7382 & 0.8663 & 0.8329 & 0.9671 \\
		Quo.Vadis~\cite{quoVadis} & 0.7741 & 0.8839 & 0.852 & 0.97 \\
		Nebula (BPE) & \textbf{0.7851} & \textbf{0.889} & \textbf{0.8548} & \textbf{0.9703} \\
		\midrule
		\multicolumn{5}{c}{Coinminer} \\
		\midrule
		Gated CNN~\cite{zhang_dmds} & 0.1066 & 0.7853 & 0.5586 & 0.9004 \\
		Neurlux~\cite{neurlux} & \textbf{0.4913} & \textbf{0.7858} & \textbf{0.691} & \textbf{0.9495} \\
		Quo.Vadis~\cite{quoVadis} & 0.083 & 0.724 & 0.4884 & 0.8964 \\
		Nebula (BPE) & 0.2468 & 0.7705 & 0.6303 & 0.9355 \\
		\midrule
		\multicolumn{5}{c}{Dropper} \\
		\midrule
		Gated CNN~\cite{zhang_dmds} & 0.1063 & 0.6901 & 0.2015 & 0.9517 \\
		Neurlux~\cite{neurlux} & \textbf{0.4433} & \textbf{0.8294} & \textbf{0.4488} & 0.9669 \\
		Quo.Vadis~\cite{quoVadis} & 0.2299 & 0.7948 & 0.358 & 0.968 \\
		Nebula (BPE) & 0.1846 & 0.7097 & 0.285 & \textbf{0.9673} \\
		\midrule
		\multicolumn{5}{c}{Keylogger} \\
		\midrule
		Gated CNN~\cite{zhang_dmds} & \textbf{0.0527} & 0.5215 & 0.0794 & \textbf{0.9418} \\
		Neurlux~\cite{neurlux} & 0.0519 & 0.5699 & 0.2032 & 0.9189 \\
		Quo.Vadis~\cite{quoVadis} & 0.0388 & \textbf{0.5848} & \textbf{0.2119} & 0.8981 \\
		Nebula (BPE) & 0.0306 & 0.5362 & 0.1295 & 0.9124 \\
		\midrule
		\multicolumn{5}{c}{Ransomware} \\
		\midrule
		Gated CNN~\cite{zhang_dmds} & 0.2318 & 0.6112 & 0.3584 & 0.899 \\
		Neurlux~\cite{neurlux} & 0.3969 & 0.6949 & 0.5527 & 0.923 \\
		Quo.Vadis~\cite{quoVadis} & 0.5385 & 0.7659 & 0.6861 & 0.9398 \\
		Nebula (BPE) & \textbf{0.6154} & \textbf{0.804} & \textbf{0.7421} & \textbf{0.9476} \\
		\midrule
		\multicolumn{5}{c}{RAT} \\
		\midrule
		Gated CNN~\cite{zhang_dmds} & 0.0099 & 0.4999 & 0.0 & 0.9276 \\
		Neurlux~\cite{neurlux} & \textbf{0.5044} & \textbf{0.7495} & \textbf{0.6625} & \textbf{0.9632} \\
		Quo.Vadis~\cite{quoVadis} & 0.0598 & 0.5326 & 0.1195 & 0.9245 \\
		Nebula (BPE) & 0.2425 & 0.6171 & 0.3683 & 0.9426 \\
		\midrule
		\multicolumn{5}{c}{Trojan} \\
		\midrule
		Gated CNN~\cite{zhang_dmds} & 0.1483 & 0.7705 & 0.5282 & 0.9355 \\
		Neurlux~\cite{neurlux} & \textbf{0.3087} & 0.8254 & 0.6153 & 0.9426 \\
		Quo.Vadis~\cite{quoVadis} & 0.1656 & 0.7674 & 0.5359 & 0.9386 \\
		Nebula (BPE) & 0.2413 & \textbf{0.8688} & \textbf{0.6827} & \textbf{0.9551} \\
            \bottomrule
        \end{tabular}
    \end{subtable}
\end{table*}

\begin{table*}[!htb]
    \caption{Results of Multi-label classification over 10 labels on Avast-CTU dataset \cite{avast_data}. Reported TPR is at FPR$=10^{-3}$.}
    \label{tab:malware_classification_avast}
    \begin{subtable}{.5\linewidth}
      \centering
        \caption{Validation set}
        \begin{tabular}{lcccc}
            \toprule
            Model & TPR & AUC & F1 & Accuracy \\
            \midrule
            		\multicolumn{5}{c}{Adload} \\
		\midrule
		Neurlux~\cite{neurlux} & \textbf{0.9927} & \textbf{0.9963} & \textbf{0.995} & \textbf{0.9998} \\
		Nebula (BPE) & 0.9686 & 0.984 & 0.9685 & 0.9988 \\
		Nebula (whitesp.) & 0.9768 & 0.9881 & 0.9738 & 0.999 \\
		\midrule
		\multicolumn{5}{c}{Emotet} \\
		\midrule
		Neurlux~\cite{neurlux} & \textbf{0.7329} & \textbf{0.9968} & \textbf{0.9957} & \textbf{0.9975} \\
		Nebula (BPE) & 0.4572 & 0.9962 & 0.9936 & 0.9962 \\
		Nebula (whitesp.) & 0.2601 & 0.9958 & 0.9929 & 0.9959 \\
		\midrule
		\multicolumn{5}{c}{Harhar} \\
		\midrule
		Neurlux~\cite{neurlux} & \textbf{1.0} & \textbf{1.0} & \textbf{0.9975} & \textbf{0.9999} \\
		Nebula (BPE) & 0.9965 & 0.9982 & 0.995 & 0.9998 \\
		Nebula (whitesp.) & 0.9965 & 0.9982 & 0.9958 & 0.9999 \\
		\midrule
		\multicolumn{5}{c}{Lokibot} \\
		\midrule
		Neurlux~\cite{neurlux} & \textbf{0.529} & \textbf{0.981} & \textbf{0.9727} & \textbf{0.9949} \\
		Nebula (BPE) & 0.3231 & 0.9729 & 0.9565 & 0.9919 \\
		Nebula (whitesp.) & 0.2655 & 0.9781 & 0.9595 & 0.9924 \\
		\midrule
		\multicolumn{5}{c}{Qakbot} \\
		\midrule
		Neurlux~\cite{neurlux} & \textbf{0.9946} & \textbf{0.9971} & \textbf{0.9956} & \textbf{0.999} \\
		Nebula (BPE) & 0.9869 & 0.9932 & 0.992 & 0.9981 \\
		Nebula (whitesp.) & 0.8387 & 0.9934 & 0.9906 & 0.9977 \\
		\midrule
		\multicolumn{5}{c}{Swisyn} \\
		\midrule
		Neurlux~\cite{neurlux} & \textbf{0.999} & \textbf{0.9994} & \textbf{0.9993} & \textbf{0.9997} \\
		Nebula (BPE) & 0.9972 & 0.9985 & 0.9983 & 0.9993 \\
		Nebula (whitesp.) & 0.9978 & 0.9989 & 0.9988 & 0.9995 \\
		\midrule
		\multicolumn{5}{c}{Trickbot} \\
		\midrule
		Neurlux~\cite{neurlux} & 0.6086 & \textbf{0.9954} & \textbf{0.986} & \textbf{0.9978} \\
		Nebula (BPE) & \textbf{0.662} & 0.9913 & 0.9825 & 0.9973 \\
		Nebula (whitesp.) & 0.5446 & 0.9928 & 0.9813 & 0.9971 \\
		\midrule
		\multicolumn{5}{c}{Ursnif} \\
		\midrule
		Neurlux~\cite{neurlux} & \textbf{0.9524} & \textbf{0.9761} & \textbf{0.9698} & \textbf{0.9987} \\
		Nebula (BPE) & 0.7346 & 0.9435 & 0.918 & 0.9964 \\
		Nebula (whitesp.) & 0.857 & 0.9562 & 0.9407 & 0.9974 \\
		\midrule
		\multicolumn{5}{c}{Zeus} \\
		\midrule
		Neurlux~\cite{neurlux} & \textbf{0.4228} & \textbf{0.9817} & \textbf{0.9663} & \textbf{0.9956} \\
		Nebula (BPE) & 0.3593 & 0.9638 & 0.9411 & 0.9923 \\
		Nebula (whitesp.) & 0.4016 & 0.9663 & 0.9474 & 0.9932 \\
		\midrule
		\multicolumn{5}{c}{Njrat} \\
		\midrule
		Neurlux~\cite{neurlux} & \textbf{0.233} & \textbf{0.9858} & \textbf{0.9606} & \textbf{0.994} \\
		Nebula (BPE) & 0.1347 & 0.975 & 0.9329 & 0.9896 \\
		Nebula (whitesp.) & 0.2328 & 0.9727 & 0.945 & 0.9916 \\
            \bottomrule
        \end{tabular}
    \end{subtable} 
    \begin{subtable}{.5\linewidth}
      \centering
        \caption{Test set}
        \begin{tabular}{lcccc}
            \toprule
            Model & TPR & AUC & F1 & Accuracy \\
            \midrule
            		\multicolumn{5}{c}{Adload} \\
		\midrule
		Neurlux~\cite{neurlux} & 0.7314 & \textbf{0.9991} & \textbf{0.715} & 0.9983 \\
		Nebula (BPE) & 0.5447 & 0.866 & 0.439 & 0.9985 \\
		Nebula (whitesp.) & \textbf{0.8248} & 0.9664 & 0.6975 & \textbf{0.9994} \\
		\midrule
		\multicolumn{5}{c}{Emotet} \\
		\midrule
		Neurlux~\cite{neurlux} & \textbf{0.7244} & 0.9345 & 0.9294 & 0.9593 \\
		Nebula (BPE) & 0.1806 & \textbf{0.9453} & \textbf{0.9392} & \textbf{0.9643} \\
		Nebula (whitesp.) & 0.4537 & 0.9381 & 0.9319 & 0.9604 \\
		\midrule
		\multicolumn{5}{c}{Harhar} \\
		\midrule
		Neurlux~\cite{neurlux} & \textbf{0.8697} & \textbf{0.9347} & \textbf{0.9031} & \textbf{0.999} \\
		Nebula (BPE) & 0.667 & 0.8333 & 0.7763 & 0.998 \\
		Nebula (whitesp.) & 0.7394 & 0.8696 & 0.8363 & 0.9984 \\
		\midrule
		\multicolumn{5}{c}{Lokibot} \\
		\midrule
		Neurlux~\cite{neurlux} & 0.0697 & 0.943 & 0.832 & 0.9777 \\
		Nebula (BPE) & 0.1583 & 0.9499 & 0.8957 & 0.9872 \\
		Nebula (whitesp.) & \textbf{0.2802} & \textbf{0.9561} & \textbf{0.9048} & \textbf{0.9882} \\
		\midrule
		\multicolumn{5}{c}{Qakbot} \\
		\midrule
		Neurlux~\cite{neurlux} & 0.4172 & 0.9945 & 0.932 & 0.9949 \\
		Nebula (BPE) & \textbf{0.9839} & 0.9918 & \textbf{0.9876} & \textbf{0.9992} \\
		Nebula (whitesp.) & 0.7505 & \textbf{0.9955} & 0.9768 & 0.9984 \\
		\midrule
		\multicolumn{5}{c}{Swisyn} \\
		\midrule
		Neurlux~\cite{neurlux} & \textbf{0.9987} & \textbf{0.9992} & \textbf{0.9991} & \textbf{0.9993} \\
		Nebula (BPE) & 0.7396 & 0.9978 & 0.9973 & 0.998 \\
		Nebula (whitesp.) & 0.9972 & 0.9985 & 0.9984 & 0.9988 \\
		\midrule
		\multicolumn{5}{c}{Trickbot} \\
		\midrule
		Neurlux~\cite{neurlux} & \textbf{0.1137} & \textbf{0.9811} & \textbf{0.9536} & \textbf{0.989} \\
		Nebula (BPE) & 0.0573 & 0.9749 & 0.9227 & 0.9811 \\
		Nebula (whitesp.) & 0.0389 & \textbf{0.9811} & 0.9056 & 0.976 \\
		\midrule
		\multicolumn{5}{c}{Ursnif} \\
		\midrule
		Neurlux~\cite{neurlux} & \textbf{0.8287} & 0.914 & 0.891 & 0.9918 \\
		Nebula (BPE) & 0.7307 & 0.9532 & 0.9362 & 0.9947 \\
		Nebula (whitesp.) & 0.7875 & \textbf{0.9684} & \textbf{0.9585} & \textbf{0.9965} \\
		\midrule
		\multicolumn{5}{c}{Zeus} \\
		\midrule
		Neurlux~\cite{neurlux} & 0.0811 & \textbf{0.9537} & 0.6503 & 0.9875 \\
		Nebula (BPE) & \textbf{0.1007} & 0.917 & 0.6419 & 0.9873 \\
		Nebula (whitesp.) & 0.1 & 0.9192 & \textbf{0.669} & \textbf{0.9894} \\
		\midrule
		\multicolumn{5}{c}{Njrat} \\
		\midrule
		Neurlux~\cite{neurlux} & 0.0633 & \textbf{0.9827} & 0.8479 & 0.9831 \\
		Nebula (BPE) & 0.078 & 0.9745 & 0.8656 & 0.9858 \\
		Nebula (whitesp.) & \textbf{0.1263} & 0.9716 & \textbf{0.8896} & \textbf{0.9887} \\
            \bottomrule
        \end{tabular}
    \end{subtable} 
\end{table*}

\end{document}

%% file: commands.tex
\newcommand{\cmark}{\ding{51}}%
\newcommand{\xmark}{\ding{55}}%

\newcommand{\vct}[1]{\ensuremath{\boldsymbol{#1}}}

\newcommand{\myparagraph}[1]{\noindent\textbf{#1}}

%% file: table1.tex
\begin{minipage}{\textwidth}
\renewcommand*\footnoterule{}
\resizebox{\textwidth}{!}{%

\begin{tabular}{cC{3cm}C{3.6cm}C{3.5cm}C{3cm}cc}
\toprule
                                     & \textbf{Data Cleaning} $\psi(\vct z)$ & \textbf{Feature Extraction} $\phi(\vct z^\prime)$ & \textbf{Model} $f(\vct x)$ & \textbf{Size} & \textbf{Code Released} & \textbf{Comment}                 \\

\midrule
Neurlux~\cite{neurlux}               & \xmark                          & Tokenization                                             & CNN, LSTM, Attention                    & 2.8M          & \cmark            &                      \\

Gated CNN~\cite{zhang_dmds}          & API filter                             & Feature Hashing                                                & CNN, LSTM                      & 0.4M          & $\sim$ & Shared privately \\

Quo.Vadis~\cite{quoVadis}            & API filter                             & Tokenization                             & CNN                               & 1.4M          & \cmark      &                           \\

JSONGrinder~\cite{avast_data} & \xmark                                & HMIL~\cite{jsonGrinder_Mandlik}                             & MLP                               & 2.4M          & $\sim$ & Non-functional   \\

CruParamer \cite{cruparamer} & API filter                          & API ``labeling''                                                & CNN, LSTM                      & --            & \xmark    &                             \\

\midrule
\textbf{Nebula (ours)}               & API, network, file, registry filters and normalization & Tokenization                                             & Transformer (Self-Attention)                 & 5.6M          & \cmark  &                               \\ 

\bottomrule
\end{tabular}
}
\end{minipage}

    




    
    
    


%% file: table2.tex
\begin{minipage}{\textwidth}
\renewcommand*\footnoterule{}
\resizebox{\textwidth}{!}{%

\begin{tabular}{ccccccccc}
\toprule
\multicolumn{2}{c}{\textbf{Malware Detection}} & \multicolumn{7}{c}{\textbf{Malware Classification}} \\
\cmidrule(lr){0-1} \cmidrule(lr){3-9}
\textbf{Benignware} & \textbf{Malware} & \textbf{Backdoor} & \textbf{Coinminer} & \textbf{Dropper} & \textbf{Keylogger} & \textbf{Ransomware} & \textbf{RAT} & \textbf{Trojan} \\
\midrule

0x406018 (1.00) & \textbf{advapi32} (1.00) & 0x1610e (1.00) & \textbf{0xcf0000} (1.00) & \textbf{0xcf0000} (1.00) & 0x48e000 (1.00) & 0x4013a0 (1.00) & \textbf{0xcf0000} (1.00) & 0x41b000 (1.00) \\

0xffe2 (0.90) & readfile (0.65) & 0x43204c (0.61) & 0x413f64 (0.79) & 0x59f934 (0.81) & 0x599c24 (0.90) & 0x42cb3a (0.71) & 0x406018 (0.78) & 0xc2b0e (0.92) \\

0x1211f9c (0.87) & 0x1211fd8 (0.07) & 0x41c024 (0.60) & 0x428084 (0.75) & 0x64 (0.81) & \textbf{0xcf0000} (0.89) & 0x4585c8 (0.71) & 0x53b9f8 (0.59) & \textbf{0xcf0000} (0.87) \\

0x481488 (0.85) & 0x1211f20 (0.06) & 0xffe2 (0.54) & \textbf{advapi32} (0.63) & 0x4635cc (0.75) & 0x5f3c24 (0.80) & 0xffe2 (0.54) & 0x42a4f1 (0.55) & \textbf{getsockobj} (0.78) \\

0x415000 (0.81) & 0x78 (0.05) & 0x53b9f8 (0.50) & 0x4130d4 (0.57) & 0x404008 (0.70) & \textbf{getsockobj} (0.67) & 0x402378 (0.51) & 0x42a730 (0.54) & 0xa6ee60 (0.77) \\

findatoma (0.76) & 0x1211f7c (0.02) & 0x414004 (0.45) & 0x42a730 (0.57) & 0x405004 (0.62) & 0x1211efc (0.62) & 0x40a175 (0.48) & 0xffe2 (0.52) & 0x404008 (0.61) \\

getcurthread (0.76) & heapalloc (0.01) & 0x425363 (0.30) & 0x402378 (0.44) & \textbf{getsockobj} (0.55) & 0x12f000 (0.57) & 0x7340 (0.47) & 0x402378 (0.46) & 123 (0.51) \\

0x40b010 (0.75) & 0x1db10106 (0.01) & 0x101c (0.28) & 0x6400000 (0.42) & 0x402566 (0.53) & 0xde10e (0.55) & \textbf{setenvnvar} (0.43) & \textbf{setenvnvar} (0.44) & 0xffe2 (0.45) \\
0x414000 (0.75) & kernel32 (0.01) & cb (0.28) & \textbf{setenvnvar} (0.42) & 0x503008 (0.45) & 0xffe2 (0.52) & 0xbbc (0.41) & 0x7090 (0.39) & 0x113000 (0.43) \\

0x7090 (0.75) & getprocheap (0.005) & 0xf0c (0.25) & 0x414c34 (0.39) & 0x408838 (0.41) & 0xfeedf030 (0.52) & 0x40c7d1 (0.41) & 0x49e5cc (0.33) & 0x80000 (0.41) \\

\bottomrule

\end{tabular}
}
\end{minipage}